\def\sbm{{\rm erg \, cm^{-2} \, s^{-1} \, arcmin^{-2} }}

\def\etal{{et al. }}

\def\cm{{\rm cm}}

\def\keV{{\rm keV}}

\documentclass{aastex}
\usepackage{emulateapj5}
\begin{document}
\lefthead{VOIT ET AL.}
\righthead{MODIFIED-ENTROPY MODELS}
%\shorttitle{MODIFIED-ENTROPY MODELS}
%\shortauthors{VOIT \& ???}
\slugcomment{\apj , received 26 February 2002, accepted 15 May 2002}
\title{Modified-Entropy Models for the Intracluster Medium}
\author{G. Mark Voit\altaffilmark{1},
        Greg L. Bryan\altaffilmark{2},
        Michael L. Balogh\altaffilmark{3},
        Richard G. Bower\altaffilmark{3}
         } 
\altaffiltext{1}{Space Telescope Science Institute,
                 3700 San Martin Drive, 
                 Baltimore, MD 21218, 
                 voit@stsci.edu}
\altaffiltext{2}{Physics Department,
                 University of Oxford, 
                 Keble Road, 
                 Oxford OX1 3RH, UK, 
                 gbryan@astro.ox.ac.uk}
\altaffiltext{3}{Department of Physics,
                 University of Durham, 
                 South Road,
                 Durham DH1 3LE, UK, 
                 M.L.Balogh@durham.ac.uk, 
                 R.G.Bower@durham.ac.uk}

\setcounter{footnote}{0}

\begin{abstract}
We present a set of cluster models that link the present-day properties of
clusters to the processes that govern galaxy formation.  These models treat the 
entropy distribution of the intracluster medium as its most fundamental  
property.  Because convection strives to establish an entropy gradient that 
rises with radius, the observable properties of a relaxed cluster depend 
entirely on its dark-matter potential and the entropy distribution of its 
uncondensed gas.  Guided by simulations, we compute the intracluster entropy 
distribution that arises in the absence of radiative cooling and supernova 
heating by assuming that the gas-density distribution would be identical to that of the dark matter.  The lowest-entropy gas would then fall below a critical 
entropy threshold at which the cooling time equals a Hubble time.  Radiative 
cooling and whatever feedback is associated with it must modify the entropy of 
that low-entropy gas, changing the overall entropy distribution function and 
thereby altering the observable properties of the cluster.  Using some 
phenomenological prescriptions for entropy modification based on the existence 
of this cooling threshold, we construct a remarkably realistic set of cluster 
models.  The surface-brightness profiles, mass-temperature relation, and 
luminosity-temperature relation of observed clusters all naturally emerge from 
these models.  By introducing a single adjustable parameter related to the 
amount of intracluster gas that can cool within a Hubble time, we can also 
reproduce the observed temperature gradients of clusters and the deviations of 
cooling-flow clusters from the standard luminosity-temperature relation. 
\end{abstract}


\section{Introduction}

The hot gas in clusters of galaxies seems like it should be easy to
understand.  Because of the relatively low ratio of baryons to dark
matter, the potential well of a cluster should be dark-matter dominated. 
The dynamical time within a cluster potential
is shorter than a Hubble time, so most clusters should be relaxed.  
Also, the cooling time of the vast majority of intracluster gas
is longer than a Hubble time.  It would appear that cluster
structure ought to be scale-free, as long as the shape of a cluster's
potential well does not depend systematically on its mass.  If that
were the case, then the global properties of clusters, such as
halo mass, luminosity-weighted temperature, and X-ray luminosity, would
scale self-similarly (Kaiser 1986).  In particular, the gas temperature 
would scale with cluster mass as $T \propto M^{2/3}$ and the bolometric 
X-ray luminosity would scale with temperature as $L \propto T^2$ in the 
bremsstrahlung-dominated regime above $\sim 2 \, \keV$.  Indeed, 
numerical simulations that include gas dynamics but exclude 
non-gravitational processes such as radiative cooling and supernova
heating produce clusters that obey these scaling laws (e.g., Evrard,
Metzler, \& Navarro 1996; Bryan \& Norman 1998; Thomas \etal 2001b).

Real clusters are not so simple.  We have known for a decade that 
the observed luminosity-temperature relation is closer to $L \propto T^3$ 
(e.g., Edge \& Stewart 1991), indicating that non-gravitational processes
must influence the density structure of a cluster's core, where most
of the luminosity is generated (Kaiser 1991; Evrard \& Henry 1991; 
Navarro, Frenk \& White 1995; Bryan \& Norman 1998).  The mass-temperature
relation, on the other hand, seemed like it ought to be more fundamental 
and less sensitive to non-gravitational effects.
Yet, observations collected over the last few years indicate that
this relation also disagrees with both the scale-free predictions 
and simulations that exclude non-gravitational processes (Horner, 
Mushotzky, \& Scharf 1999; Nevalainen, Markevitch, \& Forman 2000; 
Finoguenov, Reiprich, \& B\"ohringer 2001; Xu, Jin, \& Wu 2001).  These 
results derive mostly from resolved X-ray and temperature
profiles coupled with the assumption of hydrostatic equilibrium, 
but they do seem consistent with gravitational lensing measurements 
(Allen, Schmidt, \& Fabian 2001).

Understanding the scaling properties of clusters is of broad
importance because these scaling laws are integral to
determination of cosmological parameters.  For example, the number 
density of clusters in the local universe reflects the amplitude of
matter perturbations on $\sim 20$ Mpc scales (e.g., Henry \& Arnaud 1991;
White, Efstathiou, \& Frenk 1993; Viana \& Liddle 1996; Eke, Cole \& Frenk
1996; Kitayama \& Suto 1997; Oukbir \& Blanchard 1997; Henry 2000).
In order to measure this amplitude, one needs accurately measured
cluster masses.  Gravitational lensing measurements of cluster mass 
have grown rapidly in sophistication during the last few years, but 
X-ray determined cluster temperatures, converted to cluster masses
using the mass-temperature relation, remain the most commonly used
mass measurement for cosmological studies.  Thus, any inaccuracies 
in the mass-temperature relation propagate into uncertainties in 
cosmological parameters derived from clusters (e.g., Voit 2000).   
A number of authors have recently re-evaluated 
the perturbation amplitude implied by
cluster observations using a variety of techniques that circumvent 
the standard mass-temperature relation (Reiprich \& B\"ohringer 2001; 
Van Waerbeke \etal 2001; Seljak 2002; Viana, Nichol, \& Liddle 2002). 
The results are disconcertingly different from those based on the
standard relation, generally implying a much lower power-spectrum 
amplitude.  Clearly a better understanding of clusters is needed,
and ideally, we would like to know how the present-day properties
of clusters are linked to the physics of galaxy formation.

The physical processes most likely to break the expected self-similarity 
of clusters are heating by supernovae or active galactic 
nuclei and radiative cooling, although other possibilities 
such as magnetic pressure or cosmic-ray pressure have not been ruled 
out.  A large number of studies have investigated how the structure 
of the intracluster medium changes when these processes are included.  
Because treating all of these processes in sufficient detail is 
extremely complicated, various approximations have been employed.  
These include numerical simulations that administer a particular 
amount of energy input at early times or adopt some other 
phenomenological prescription for injecting energy
(e.g., Metzler \& Evrard 1994; Navarro, Frenk, \& White 1995; 
Metzler \& Evrard 1997; Bialek, Evrard, \& Mohr 2001; Borgani et al. 2001a; 
da Silva \etal 2001; Bryan \& Voit 2001), analytical models for
spherically symmetric accretion of preheated gas (e.g., Cavaliere, Menci
\& Tozzi 1999; Tozzi \& Norman 2001), and semi-analytic models that
consider the merger and star formation history of clusters (e.g., Wu, Fabian,
\& Nulsen 1998, 2000; Balogh, Babul, \& Patton 1999; Bower \etal 2001;
Babul, Balogh, Lewis, \& Poole 2002).  Most of these papers have
focused on heating of the gas, but recent work has also examined the
possibility that radiative cooling is primarily responsible 
(e.g., Knight \& Ponman 1997; Suginohara \& Ostriker 1998; 
Pearce \etal 2000; Bryan 2000; Muanwong \etal 2001;
Thomas \etal 2001a; Lewis \etal 2001).  Two broad conclusions emerge
from such studies: (1) radiative cooling without feedback probably 
locks too high a fraction of baryons into condensed objects, and (2)
if heating is uniformly distributed, the amount required 
to produce realistic clusters is quite large, probably exceeding $\sim
1 \, \keV$ per baryon.  Introducing this much energy through
supernova heating strains the bounds of plausibility, given how many
metals are observed in the intracluster medium (e.g., Balogh, Babul, \&
Patton 1999; Valageas \& Silk 1999; Kravtsov \& Yepes 2000;  Bower
\etal 2000; although see Loewenstein 2000).

Here, we revisit an approach pioneered by Bower (1997), focusing 
on entropy as the most fundamental characteristic of the
intracluster medium.  Analyzing clusters in terms of entropy
has two advantages.  The first is that convection within clusters
acts as an entropy-sorting device, shuttling low-entropy material 
to the cluster core and high-entropy material to the cluster's outskirts.
Convective stability is achieved when specific entropy becomes a
monotonically increasing function of radius.  Thus, if we can 
determine the entropy distribution of the gas in a cluster and
we know the structure of its dark-matter halo, we can predict the structure 
of that cluster in its relaxed state.  In the idealized cases that
we will consider, we will assume that clusters are spherically 
symmetric, dark-matter dominated, in hydrostatic equilibrium,
and in convective equilibrium.  The second advantageous property
of entropy is that, in the absence of non-gravitational processes,
the entropy distribution of a cluster can be readily computed 
using either hydrodynamical simulations or spherically symmetric
analytical models.  Then, the effects of non-gravitational
heating and radiative cooling can be implemented as modifications
of that entropy distribution.

In Voit \& Bryan (2001), we introduced a simple scheme for
computing cluster models, taking advantage of these properties
of the intracluster entropy distribution to show how heating
and cooling conspire to produce a cluster $L$-$T$ relation quite 
similar to what is observed.  This paper explores those 
modified-entropy models in more detail, showing that the observable
properties of these model clusters are quite similar to those of
real clusters on many counts.  Section 2 describes how we
construct those models and analyzes how cluster properties
depend on both halo concentration and entropy modification.
Section~3 uses relations between halo mass and halo concentration
to define sets of cluster models determined entirely by the 
underlying cosmology.  We then compare the model clusters
with observations of the surface-brightness profiles of clusters,
the temperature gradients of clusters, the mass-temperature relation,
and the luminosity-temperature relation.  In every case, our models
reproduce the observations, with little need for adjustable parameters.
In section 4, we explore how the present-day intracluster entropy
distribution emerges from hierarchical cosmological models.
Section~5 summarizes our conclusions.

\section{Modified Entropy Models}

%The most fundamental property of the intracluster medium is its 
%distribution of specific entropy with mass.  If one knows both 
%the shape of the dark-matter potential that confines the cluster's
%gas and the entropy distribution function of that gas, then one 
%can calculate the observable properties of the equilibrium 
%configuration.

In this section we develop a family of modified-entropy models
of clusters based on two parameters, one specifying the concentration
of the cluster's dark-matter halo and the other specifying the
entropy level at which cooling and feedback modify the intracluster
entropy distribution.  We begin by casting the equilibrium equations
in dimensionless form and selecting appropriate boundary conditions.
Then we assume that the unmodified entropy distribution determined
by gravitational processes alone would produce a gas density distribution
similar to the dark-matter density distribution.  Because the innermost
gas in such a cluster can cool in much less than a Hubble time,
we proceed to investigate several potential modifications of the entropy
distribution by radiative cooling and subsequent feedback.  In each
case, we show how both entropy modification and the shape of the 
underlying dark-matter halo affect the surface-brightness profile 
of a cluster, its luminosity-weighted temperature, its temperature 
gradient, and its X-ray luminosity.  Two important results
emerge from this analysis: (1) the temperature of a cluster of a
given mass is determined primarily by halo concentration and is
affected only modestly by entropy modification, and (2) altering
the intracluster entropy distribution at the scale set by radiative
cooling leads to an $L_X$-$T$ relation that has the observed slope
at $z=0$ and that evolves very little with time.

\clearpage
\subsection{Equilibrium Structure}
    
The equilibrium structure of the intracluster medium is determined
by three things, the gravitational potential of the cluster's dark 
matter, the entropy distribution of the cluster's gas, and the
confining pressure at the outer boundary of the cluster.  In order
to solve for that structure, one must integrate the equations
of hydrostatic equilibrium and gas mass conservation,
\begin{eqnarray}
   \frac {dP} {dr} = - \frac {GM(<r)} {r^2} \rho  \\
   \frac {dM_g} {dr} = 4 \pi r^2 \rho \; \; ,
\end{eqnarray}
using the equation of state $P = K \rho^\gamma$, where $\gamma$ is
the adiabatic index and $K$ specifies the adiabat of the gas.
In this system of equations, $M(<r)$ is the total mass within radius 
$r$, $M_g$ is the gas mass within that radius, and the other symbols 
have their usual meanings.  Throughout this paper, we will assume
that the gas mass is gravitationally negligible and that $\gamma = 5/3$,
as appropriate for an ideal monatomic gas.  

In a sufficiently relaxed cluster, convection ensures that $K(r)$ 
monotonically increases with radius.  Thus, if one knows $M_g(K)$,
the entropy distribution of the intracluster medium expressed in
terms of the mass of gas with $P\rho^{-5/3} < K$, then one can 
use the inverse relation $K(M_g)$ to solve the equilibrium 
equations that determine the cluster's structure.  Because this
crucial quantity $K$ is so closely related to specific entropy, 
we will often refer to it as the ``entropy'' of the gas,
even though the standard thermodynamic entropy per particle
for an ideal monatomic gas is $s = \ln K^{3/2} + {\rm const.}$  

\subsubsection{Dimensionless Form}
\label{dimensionless}

We can gain insight into the key parameters that control the
structure of the intracluster medium by investigating a family 
of dimensionless models for clusters in hydrostatic and convective
equilibrium.  Because the virial radius of a cluster lies 
close to the radius $r_{200}$ 
within which the mean matter density of the cluster is 200 times 
the critical density $\rho_{cr}$, we elect to represent radii 
in terms of $\hat{r} \equiv r/r_{200}$ and gas density in terms of 
$\hat{\rho} \equiv \rho / f_b \rho_{200}$,
where $\rho_{200} = 200 \rho_{cr}$ and 
$f_b = 0.02 / \Omega_m h^2$ is the fractional contribution 
of baryons to the total mass of the universe.  
One can then rewrite the equilibrium equations in dimensionless 
form as follows:
\begin{eqnarray}
  \frac {d\hat{P}} {d\hat{r}} = - 2 \frac {\hat{M}} {\hat{r}^2} \hat{\rho} \\
  \frac {df_g} {d\hat{r}} = 3 \hat{r}^2 \hat{\rho} \; \; ,
\end{eqnarray}
where 
$M_{200} = (4 \pi / 3) r_{200}^3 \rho_{200}$ is the virial mass,
$T_{200} = G M_{200} \mu m_p / 2 r_{200}$ is the temperature
in energy units\footnote{Boltzmann's constant $k$ is absorbed into
$T$ throughout the paper.} of the corresponding singular isothermal sphere, 
$\hat{P} = P  / [T_{200} f_b \rho_{200} (\mu m_p)^{-1}]
$ is the dimensionless pressure, 
$\hat{M} = M(<r) / M_{200}$ is the dimensionless mass within $\hat{r}$,
and $f_g = M_g / (f_b M_{200})$ is the fraction of a cluster's baryons
in the intracluster medium within radius $\hat{r}$.
The corresponding dimensionless entropy distribution is 
$\hat{K}(f_g) = K / [T_{200} (f_b \rho_{200})^{-2/3}
(\mu m_p)^{-1}]$, and we will also make use of the quantity
$\hat{T} = P \rho^{-1} \mu m_p T_{200}^{-1}$.  

General solutions to these equations can be found for a few idealized 
cases.  For example, if the potential is a singular isothermal sphere 
and the gas is also isothermal, then we have $\hat{P} \propto \hat{\rho} 
\propto \hat{r}^{-2}$, $f_g \propto \hat{r}$, and 
$\hat{K} \propto \hat{r}^{4/3}$ for a monatomic ideal gas.  If instead 
the mass profile follows the form of Navarro, Frenk \& White 
(1997; NFW hereafter), with $\hat{M} \propto [ \ln (1+c\hat{r}) 
- c\hat{r} (1+c\hat{r})^{-1} ]$ 
where $c$ is the concentration parameter, then the equation of 
hydrostatic equilibrium becomes (see Wu, Fabian, \& Nulsen 2000)
\begin{equation}
   \frac {d\hat{P}} {d\hat{r}} = 2 \hat{\rho}  
                           \left[ \ln (1+c) - \frac {c} {1+c} \right]^{-1}
                           \frac {d} {d\hat{r}} 
			   \left[ \frac {\ln (1+c\hat{r})} {\hat{r}} \right] 
                           \; \; .
\end{equation}
Given an isentropic gas in which $\hat{P} \hat{\rho}^{-5/3} = \hat{K}
= {\rm const.}$ we thus obtain $\hat{\rho} \propto [ \hat{r}^{-1} 
\ln (1+c\hat{r}) + C]^{3/2}$, where $C$ is a constant of integration.

\subsubsection{Boundary Conditions}

A particular solution to the equilibrium equations can be found by
choosing $\hat{P}$ and $f_g$ at the origin and integrating outwards.
The gas mass fraction at the origin is always zero by definition,
but there is some freedom in the choice of $\hat{P}_{0} \equiv
\hat{P}(0)$.  Any given choice of $\hat{P}_{0}$ corresponds to a 
unique ICM structure with a unique luminosity and temperature
profile, but which solutions are the physical ones?

Here is where the confining pressure comes into play.  It is
generally assumed that accreting matter confines the ICM in
the neighborhood of $r_{200}$.  The pressure at $r_{200}$ 
would then be determined by the ram pressure of the infalling 
matter.  Now suppose that $K (M_g)$ has been modified by 
non-gravitational processes.  For example, heating might drive 
gas out of the cluster potential, pushing the accretion shock 
to a radius beyond $r_{200}$.  Alternatively, cooling and condensation 
might reduce the total gas mass of the intracluster medium 
so that the accretion shock moves inward.  In either
case the radius at which the integration of hydrostatic 
equilibrium should terminate differs from $r_{200}$.

So what is the termination radius $r_{max}$ at which 
$f_g f_b M_{200}$ equals the total uncondensed gas mass, 
and what is the pressure at that radius?
In a freely falling accretion flow with a constant mass flux, 
the density should vary as $v_{ff}^{-1} r^{-2}$, where 
$v_{ff} \propto \hat{r}^{-1/2}\sqrt{ \ln (1 + c\hat{r}) }$
is the free-fall velocity from $\hat{r} \gg 1$ in an NFW potential.  
We therefore assume that the accretion pressure scales with radius
as $\hat{P}_{acc}(\hat{r}) \propto \hat{r}^{-5/2} \sqrt{ \ln (1+c\hat{r})}$, 
and we normalize that pressure to equal the unmodified NFW value
(see \S~\ref{unmod}) at $\hat{r} = 1$. Various choices of $\hat{P}_0$ 
will lead to pressure profiles that intersect $\hat{P}_{acc}(\hat{r})$ 
at different values of $\hat{r}_{max}$, each corresponding to a 
unique value of $f_g(\hat{r}_{max})$ determined by a particular 
integration.  The correct solution is the one that gives the proper 
value for the uncondensed gas mass at the radius where the gas pressure 
equals the accretion pressure.

In practice, the physically reasonable solutions are not 
particularly sensitive to the outer boundary condition.  In the 
models that follow, applying the boundary condition $\hat{P}  = \hat{P}_{acc}$ at $\hat{r} = 1$ would lead to similar results.  
However, the value of $f_g$ derived at $\hat{r}=1$ using this 
boundary condition is not physically significant and can even 
be inconsistent with models in which cooling and condensation 
significantly reduce the maximum value of $f_g$.  One example 
of a physically {\em unreasonable} boundary condition is requiring
$f_g = 1$ at $\hat{r} = 1$.  Such a requirement does not allow 
intracluster gas to expand beyond $r_{200}$ even when heating 
substantially raises the intracluster entropy.  In the limit 
of extreme heating, this condition therefore artificially boosts 
the central pressure, density, and temperature, leading to an 
unphysically large cluster luminosity.

\subsubsection{Integrated Characteristics}
\label{integrated}

Once a correct solution has been identified, one can compute
two quantities that correspond to the dimensionless luminosity
and luminosity-weighted temperature:
\begin{eqnarray}
  \hat{L} & = & 3 \int_0^{\hat{r}_{max}} \hat{\rho}^2 \hat{r}^2 d\hat{r} \\
  \hat{T}_{\rm lum} & = & 3 \hat{L}^{-1} \int_0^{\hat{r}_{max}} \hat{T} 
                        \hat{\rho}^2 \hat{r}^2 d\hat{r}
               \; \; .
\end{eqnarray}
As long as the cooling function $\Lambda(T)$ remains sufficiently 
constant within the gas contributing the bulk of the luminosity, 
the actual luminosity-weighted temperature is $T_{\rm lum} \approx T_{200} 
\hat{T}_{\rm lum}$, and the bolometric X-ray luminosity is
\begin{equation}
  L \approx \frac {4 \pi} {3} r_{200}^3 
                     \left( \frac {n_p} {n_e} \right)
                     \bar{n}_e^2 
                     \Lambda(T_{\rm lum}) \hat{L} \; \; ,
\end{equation}
where $\bar{n}_e = (n_e / \rho) f_b \rho_{200} = 1.2 \times 10^{-4}
\, (\Omega_M / 0.33)^{-1} \, {\rm cm}^{-3}$.  
Scaling to a 10~keV cluster at $z=0$ gives
\begin{eqnarray}
\label{eq-lx}
L & \approx & (9.0 \times 10^{43} \, h^{-3} \, {\rm erg \, s}^{-1})
             \left( \frac {T_{200}} {10 \, \keV} \right)^{3/2}
             \left( \frac {\Omega_M} {0.33} \right)^{-2}
                \nonumber \\
      &  &     \hspace{2.0cm} \times \; 
             \Lambda_{23}(T_{\rm lum}) \hat{L} \; ,
\end{eqnarray}
where $\Lambda_{23}(T_{\rm lum}) = \Lambda(T_{\rm lum}) / (10^{-23} \, 
{\rm erg \, cm^3 \, s^{-1}})$.

The emission-measure profile and emissivity-weighted temperature
profile are also of interest.  Thus, we define the following dimensionless
analogs as functions of the projected radius $\hat{r}_\perp$:
\begin{eqnarray}
  \hat{S}(\hat{r}_\perp) & = & 
                        2 \int_0^{\sqrt{\hat{r}_{max}^2-\hat{r}_\perp^2}} 
                         \hat{\rho}^2 d\hat{l} \\
  \hat{T}_{\perp}(\hat{r}_\perp) & = &  2 \hat{S}^{-1}
                         \int_0^{\sqrt{\hat{r}_{max}^2-\hat{r}_\perp^2}} 
			  \hat{T} \hat{\rho}^2 d\hat{l} 
               \; \; ,
\end{eqnarray}
where $\hat{l} = \sqrt{\hat{r} - \hat{r}_\perp}$ is the dimensionless
distance along the line of sight.

\subsection{Unmodified Solutions}
\label{unmod}

Before exploring the consequences of modifying the entropy of
intracluster gas, we need to know what the entropy distribution
would be if it were not modified.  To simplify matters, we will assume
that the underlying dark matter profile of the cluster is of NFW form with
some concentration parameter $c$.  This parameter typically ranges
from $c \sim 5$ for hot clusters to $c \sim 10$ for groups of 
galaxies (see \S~\ref{conparam}).  If the intracluster medium 
were collisionless, then its density profile would be identical 
to that of the dark matter.  Thus, we define the unmodified
entropy distribution $K_0(M_g)$ of a cluster of concentration $c$
to be that of gas in hydrostatic equilibrium in the cluster potential
with a density profile identical to that of the dark matter.

% ------------------ fig -----------------
\vspace{\baselineskip}
%\begin{figure}
\epsfxsize=3in 
\centerline{\epsfbox{f1.epsi}}
\figcaption{\footnotesize
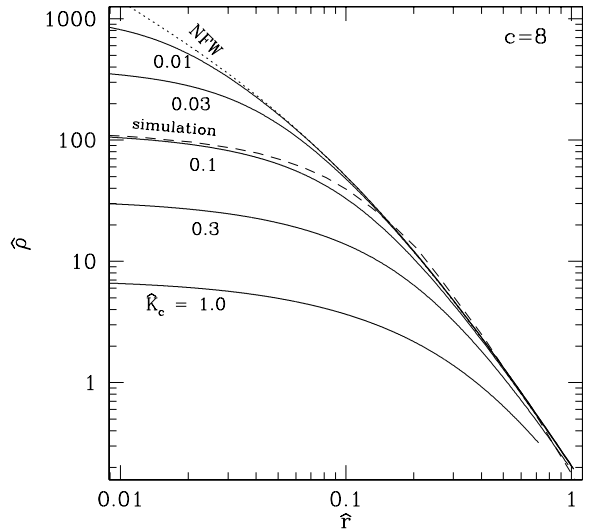
Intracluster density profiles in
dimensionless units.  The dotted line shows the matter
density ($\hat{\rho}$) of an NFW profile of concentration $c=8$ in 
units of the mean density within $r_{200}$ as a function of radius
($\hat{r}$) in units of $r_{200}$.  The dashed line shows
the density profile of a simulated cluster whose dark matter
halo is well approximated by an NFW halo with $c=8$.  Outside
of $\hat{r} = 0.1$, these curves closely correspond.  The solid
lines show how the run of density with radius changes as the
entropy distribution of an unmodified $c=8$ NFW halo is truncated
at progressively larger values of the dimensionless entropy
$\hat{K}_c$.  For these cases, $\hat{\rho}$ is the gas density
in units of the mean density of the unmodified gas.  Notice that
the model with $\hat{K}_c = 0.1$ is quite similar to the gas 
density distribution from the simulation.
\label{denprofs}}
\vspace{\baselineskip}
%\end{figure}
% ----------------------------------------

Numerical experiments reveal that the entropy distribution of 
intracluster gas in clusters simulated without radiative
cooling or supernova heating is quite similar to the NFW form
throughout most of the cluster but is elevated above NFW within 
the cluster's core.  Figure~\ref{denprofs} shows the density
distribution of a cluster simulated with an adaptive-mesh
refinement code (Norman \& Bryan 1998; Bryan 1999) 
for the Santa Barbara cluster comparision
project (Frenk \etal 1999).  The dotted line labeled ``NFW'' shows an
NFW density distribution with $c=8$, which is very close to
the underlying dark-matter distribution of the cluster.
The dashed line labeled ``simulation'' shows the gas-density
distribution from that same simulation.  Agreement is quite
close outside 10\% of the virial radius, but the gas density
levels off within that radius, implying that the lowest-entropy
gas has a somewhat higher entropy than assumed in our
unmodified distribution $K_0$.  However, this discrepancy at
the low-entropy end of the distribution is inconsequential
to the modified-entropy models that follow because the entropy
of the discrepant gas will always be subject to further 
modification.

The unmodified distribution is also somewhat discrepant with
simulations at the high-entropy end.  Assuming hydrostatic equilibrium 
and an NFW gas-density law near the virial radius leads to gas 
temperatures that are 20-30\% higher than those in our fiducial 
simulation.  Thus, our unmodified entropy levels at large radii 
are slightly higher than what a numerical simulation would produce.  
However, in order to properly reproduce the density profile near 
the virial radius in a hydrostatic model, we must retain these elevated
entropy levels.  This overestimate at the cluster's outskirts has
virtually no effect on emissivity-weighted global quantities, such
as $L$ and $T_{\rm lum}$, but if one  

\clearpage \noindent
is interested in the temperature of cluster gas near $r_{200}$,
then departures from hydrostatic equilibrium, which are outside 
the scope of our models, must be taken into account.

\subsection{Entropy Modification}

Modification of this baseline entropy distribution is inevitable 
because clusters simulated without radiative cooling are physically 
inconsistent.  In general, the gas at the center of such a simulated 
cluster can radiate many times its thermal energy within a Hubble 
time, resulting in cluster luminosities that vastly exceed those 
observed (e.g., Muanwong \etal 2001; Bryan \& Voit 2001). 
Condensation and removal of the lowest-entropy 
gas from the intracluster medium must happen at some level, because 
that is how the cluster's galaxies form stars.  Furthermore, the 
most massive of these stars must explode, resulting in supernova 
feedback.  Condensed gas accreting onto an active galactic nucleus 
can provide additional feedback (Heinz, Reynolds, \& Begelman 1998;
Kaiser \& Alexander 1999; Quilis, Bower, \& Balogh 2001; Reynolds,
Heinz, \& Begelman 2002; B\"ohringer \etal 2002).  All of these processes 
will modify the entropy distribution of the intracluster medium,
but modeling them in detail is a daunting task.

Because the physics of feedback is so complex, we shall adopt a
highly simplistic phenomenological approach to entropy
modification.  Instead of trying to model all the consequences of
heating and cooling we will restrict our investigation to three 
qualitatively different modifications of the entropy distribution 
$K_0$: (1) truncation of the distribution, (2) shifting of the 
distribution, and (3) radiative losses from the distribution.  
Each of these modifications depends on a single entropy threshold
parameter $K_c$, which we take to be the entropy 
at which the cooling time of 
gas of temperature $T_{200}$ equals the age of the universe.  
Thus, for each type of entropy modification, we obtain a two-parameter 
family of modified-entropy models specified by a concentration $c$
and a dimensionless entropy $\hat{K}_c$.

As in Voit \& Bryan (2001), we use an approximate cooling function 
for gas with a metallicity of one-third solar to determine the appropriate
entropy threshold.  According to this approximation, the threshold 
for cooling within 15~Gyr can be expressed as
\begin{equation}
\label{eq-Kc}
  K_c \approx \frac {135 \, \keV \, \cm^2} {\mu m_p}
            \,  \left( \frac {\rho} {n_e} \right)^{-2/3}
              \left( \frac {T_{\rm lum}} {2 \, \keV} \right)^\zeta  \; \; ,
\end{equation}
with $\zeta = 2/3$ for $T > 2 \, \keV$ and $\zeta = 0$ for $T <
2 \, \keV$.  Converting this threshold to dimensionless units
yields
\begin{equation}
\label{eq-hatKc}
  \hat{K}_c \approx 0.164 \: \hat{T}_{\rm lum} 
	      \, \left( \frac {T_{\rm lum}} {2 \, \keV} \right)^{\zeta-1} 
                 \left( \frac {\Omega_M} {0.33} \right)^{-2/3} \; \; .
\end{equation}
As long as $\hat{T}_{\rm lum} \approx 1$ (see \S~\ref{emodtemp}), 
then $\hat{K}_c \approx 0.1$ will be appropriate
for the hottest clusters ($\sim 10 \, \keV$) and $\hat{K} 
\approx 0.5-1.0$ will be appropriate for the coolest 
($\sim 0.5 \, \keV$) groups.

\subsubsection{Truncation}

One simple yet physically motivated way to modify the 
entropy distribution of intracluster
gas is to truncate it at $\hat{K}_c$ (Bryan 2000; Voit \& Bryan 2001).
Mathematically, we 

% ------------------ fig -----------------
\vspace{\baselineskip}
%\begin{figure}
\epsfxsize=3in 
\centerline{\epsfbox{f2.epsi}}
\figcaption{\footnotesize
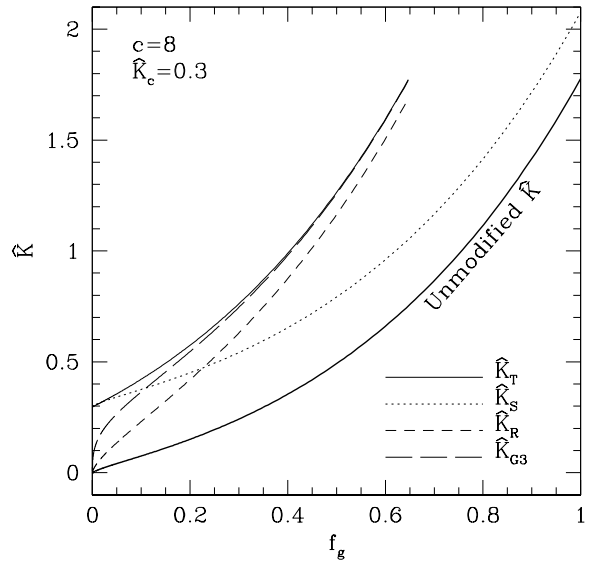
Modified entropy distributions.  The thick solid line shows how 
dimensionless entropy $\hat{K}$ rises with $f_g$, the fraction of
a cluster's baryons in the intracluster medium within radius $\hat{r}$,
for an unmodified $c=8$ NFW halo.  Gas in the unmodified halo is assumed 
to be in hydrostatic equilibrium with a density profile identical to 
that of the dark matter.  The thin solid line illustrates the entropy
distribution $\hat{K}_{\rm T}(f_g)$ that results when the unmodified 
distribution is truncated at $\hat{K}_c = 0.3$, corresponding to a
horizontal translation of the unmodified distribution.   The dotted line
shows $\hat{K}_{\rm S}(f_g)$, the unmodified distribution shifted vertically
by $\hat{K}_c = 0.3$.  The long- and short-dashed lines show two
different modifications designed to mimic radiative losses from
the unmodified distribution (see \S~\ref{radloss}).  These lines
correspond to a reduction of $\hat{K}$ at each point of the unmodified
distribution, followed by a horizontal translation so that 
$\hat{K}(0) = 0$.
\label{kdist}}
\vspace{\baselineskip}
%\end{figure}
% ----------------------------------------

\noindent
define the truncated distribution to be
\begin{equation}
 \hat{K}_{\rm T}(f_g) \equiv \hat{K}_0(f_g + f_c) \; \; ,
\end{equation}
where $f_c$ is defined by $\hat{K}_0(f_c) = \hat{K}_c$.
Such a modification corresponds to removing all the gas with a cooling 
time less than a Hubble time and ignoring the effects of cooling on 
the rest of the intracluster medium.  
This type of modification can viewed either
as an extreme form of cooling, in which all the gas below the critical
threshold condenses, or an extreme form of heating, in which all
the gas below the threshold is heated to very high entropy ($\hat{K}
\gg 1$) and convects beyond the virial radius of the cluster.  
Figure~\ref{denprofs} shows some gas density profiles resulting
from different values of $\hat{K}_c$.  The thin solid 
line labeled $\hat{K}_{\rm T}$ in Figure~\ref{kdist} shows the truncated 
entropy distribution for $c = 8$ and $\hat{K}_c = 0.3$. 

Notice that the maximum value of $f_g$ in these truncated models
is less than unity because some of the gas has been removed
from the intracluster medium.  The thick solid line in Figure~\ref{rmax} 
illustrates how the fraction $f_{g,max} = 1 - f_c$ of a cluster's baryons that
remain in the intracluster medium depends on $\hat{K}_c$.  For hot 
clusters this fraction is $\sim$90\%, but for cool groups 
it drops below 50\%.  Nevertheless, the outer radius $\hat{r}_{max}$
of the intracluster gas remains close to $r_{200}$, even when 
$f_{g,max}$ drops below 0.5.  Figure~\ref{denprofs} shows the reason.
As the entropy threshold for truncation rises, the central 
entropy of the cluster also rises.  The resulting density profile is
therefore much flatter for large values of $\hat{K}$, 
enabling a smaller amount of gas \\

% ------------------ fig -----------------
\vspace{\baselineskip}
%\begin{figure}
\epsfxsize=3in 
\centerline{\epsfbox{f3.epsi}}
\figcaption{\footnotesize
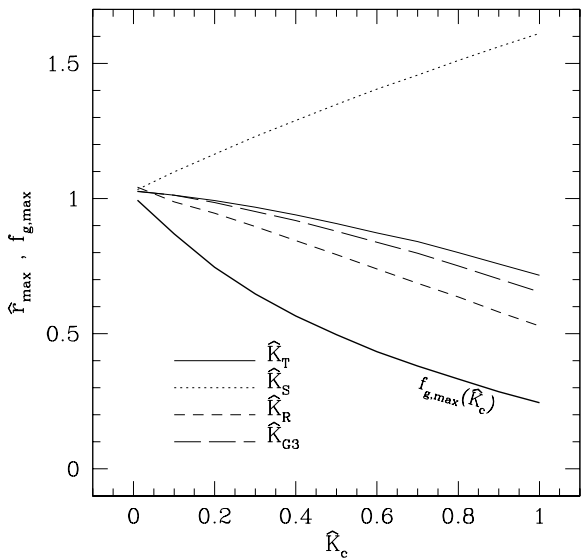
Maximum radius and baryon content of the intracluster medium in 
modified-entropy models.  The thick solid line shows how
$f_{g,max}$, the fraction of a cluster's baryons residing
in the intracluster medium, depends on the entropy threshold
$\hat{K}_c$ in the modified distributions $\hat{K}_{\rm T}$,
$\hat{K}_{\rm R}$, and $\hat{K}_{\rm G3}$.  The other lines
show how $\hat{r}_{\rm max}$, the maximum radius of the intracluster
medium, depends on $\hat{K}_c$ in each of these modified
distributions.
\label{rmax}}
\vspace{\baselineskip}
%\end{figure}
% ----------------------------------------

\noindent
to fill a similar volume.  Note also that the 
density profile of a modified-entropy model with $c = 8$ and $\hat{K}_c
= 0.1$ is quite similar to that of a $c=8$ cluster simulated without
cooling and feedback, suggesting that the cooling threshold may have 
only a modest effect on the structure of the hottest clusters.

\subsubsection{Shift}

Another way to modify intracluster entropy is to add a constant
term $\hat{K}_c$ to the unmodified entropy distribution (Voit \& Bryan
2001).  This type of modification is a convenient way to mimic
``preheated'' models for cluster formation that inject a 
fixed amount of entropy per particle into the intergalactic 
medium at some early time.  In this case, the
modified entropy distribution is defined by
\begin{equation}
 \hat{K}_{\rm S}(f_g) \equiv \hat{K}_0(f_g) + \hat{K}_c \; \; .
\end{equation}
The dotted line in Figure~\ref{kdist} shows this distribution for
$c=8$ and $\hat{K}_c = 0.3$.  Because entropy has been added but 
no gas has been removed, the outer radius for models with a shifted 
entropy distribution extends beyond the usual virial radius (see
Figure~\ref{rmax}).  Qualitatively, this corresponds to heating
that inflates the intracluster medium, driving the accretion
shock to larger radii.

Comparing the structure of shifted models to truncated
models reveals interesting similarities.  Figure
\ref{denprofs_all} depicts the dimensionless density profiles 
for $\hat{K}_c = 0.1$, 0.3, and 1.0 in clusters with $c=8$.
The run of density with radius is virtually identical for the
shifted and truncated models when $\hat{K}_c = 0.1$ and is still 
quite similar when $\hat{K}_c = 0.3$.  Thus, for most
of the interesting range of entropy thresholds, there is
little observable difference between shifted models and
truncated models.  However, as $\hat{K}_c$ rises to 1.0, 
the density profile of the shifted model becomes noticeably 
flatter than that of the truncated model.

% ------------------ fig -----------------
\vspace{\baselineskip}
%\begin{figure}
\epsfxsize=3in 
\centerline{\epsfbox{f4.epsi}}
\figcaption{\footnotesize
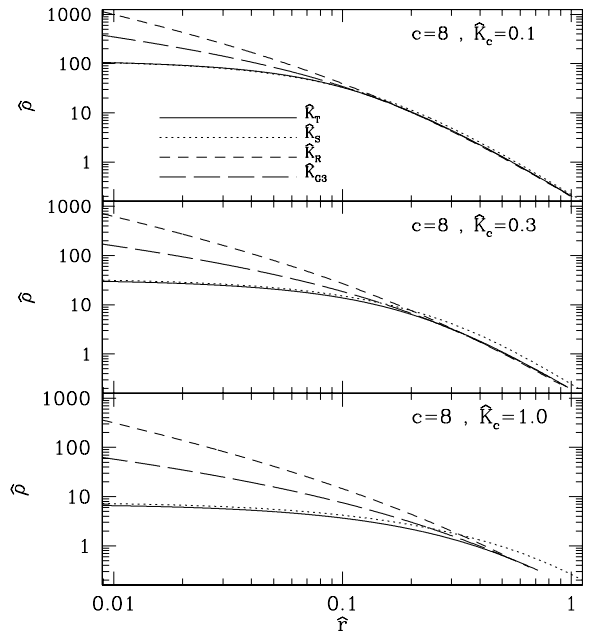
Density profiles for modified
entropy models with $c=8$ and $\hat{K}_c = 0.1$, 0.3, and
1.0.  Solid lines show $\hat{\rho}(\hat{r})$ for truncated
models ($\hat{K}_{\rm T}$), dotted lines represent shifted
models ($\hat{K}_{\rm S}$), and the long- and short-dashed
lines represent two types of radiative-loss model ($\hat{K}_{\rm R}$
and $\hat{K}_{\rm G3}$).  The density structure of truncated 
and shifted models is almost indistinguishable for small values
of $\hat{K}_c$.  The radiative-loss models have the elevated
central densities characteristic of cooling-flow clusters.
\label{denprofs_all}}
\vspace{\baselineskip}
%\end{figure}
% ----------------------------------------

Similarity between the truncated and shifted models at the lower
entropy thresholds arises because the unmodified entropy $\hat{K}_0$
depends approximately linearly on $f_g$ for small $\hat{K}$.
Thus, shifting the distribution by $\hat{K}_c$ and truncating
it at $\hat{K}_c$ both produce similar modified entropy distributions
and therefore similar density, temperature, and pressure 
distributions in the inner regions of the cluster.  That 
is why entropy modification by heating and cooling can have 
similar effects on the luminosity and temperature of clusters
as long as they establish similar values of the minimum
entropy.

\subsubsection{Radiative Losses}
\label{radloss}

Observations show that many clusters contain gas with a cooling 
time less than a Hubble time (Fabian 1994), motivating us 
to consider slightly more complicated modifications to
the entropy distribution.  Radiative cooling reduces the
the specific entropy of a gas parcel according to
\begin{equation}
 \frac {ds} {dt} = \frac {d \ln K^{3/2}} {dt}
                 = - \frac {\mu m_p n_p n_e \Lambda(T)} {\rho T}
                         \; \; .
\end{equation}
This entropy equation can be rewritten to eliminate the
dependence on gas density:
\begin{equation}
 \frac {d K^{3/2}} {dt} = - \left( \frac {n_p n_e} {\rho^2} \right)
                            \, (\mu m_p)^{-1/2}
                            \, T^{1/2} \Lambda(T) \; \; .
\end{equation}
In general, the rate at which entropy declines will depend
on the prevailing hydrodynamic conditions, 
standard examples being isobaric cooling or isochoric cooling.  
However, for the purposes of
this toy model, let us assume that the right-hand side of
this equation remains constant.  We make this rather arbitrary 
assumption for two physical reasons.  First, within
the temperature range $0.1 \, \keV \lesssim T \lesssim 2 \, \keV$, 
the quantity $T^{1/2} \Lambda(T)$ is approximately constant.
Second, as gas settles within a nearly isothermal potential,
work done by compression will attempt to maintain the temperature
of that gas near the characteristic temperature of the halo.
Thus, we are motivated to use the following modification to
represent radiative losses:
\begin{equation}
 \hat{K}_{\rm R}(f_g) \equiv \{ [\hat{K}_0(f_g + f_c)]^{3/2} 
                        - \hat{K}_c^{3/2} \}^{2/3} \; \; .
\end{equation}
This modification corresponds to reducing the value of
$\hat{K}^{3/2}$ by the constant amount $\hat{K}_c^{3/2}$ throughout the
entropy distribution and discarding all the gas with $\hat{K} < 0$.
The short-dashed line in Figure~\ref{kdist} 
shows an example of the resulting distribution for $c=8$ and 
$\hat{K}_c = 0.3$, and a similar line in Figure~\ref{rmax} 
shows how the maximum radius depends on $\hat{K}_c$.

It will also be interesting to investigate a generalization of
this radiative-loss model:
\begin{equation}
 \hat{K}_{{\rm G}\alpha}(f_g) \equiv \{ [\hat{K}_0(f_g + f_c)]^{\alpha} 
                        - \hat{K}_c^{\alpha} \}^{1/\alpha} \; \; ,
\end{equation}
in which the tunable parameter $\alpha$ enables us to adjust the amount 
of gas below the cooling threshold.  In the limit $\alpha 
\rightarrow \infty$, this generalized distribution becomes 
identical to the truncated distribution.  In what follows, 
we will focus on the intermediate case of $\alpha = 3$, represented 
by the symbol $\hat{K}_{\rm G3}$ and long-dashed lines in 
the relevant figures.

Figure~\ref{denprofs_all} compares the density distributions that
result from the radiative-loss models with the truncated and
shifted models.  For $\hat{K}_c = 0.1$, all the density profiles
are very similar at $\hat{r} > 0.1$, because the 
entropy profiles are nearly identical there.  However, within
that radius, entropy in the radiative-loss models drops below
$\hat{K}_c$, allowing for considerably greater compression and
higher central density.  This trend becomes more pronounced at higher
$\hat{K}_c$ because the proportion of the intracluster medium
below the cooling threshold is even larger.

\subsection{Entropy Modification and Surface Brightness}  
\label{emodsb}

Modified-entropy models generated through shifts and truncations
of the unmodified entropy distribution produce surface-brightness
profiles that are strikingly similar to the $\beta$-model profiles
of many real clusters, at least within about 30\% of the virial
radius.  Figure~\ref{sbright} shows the dimensionless emission
measure $\hat{S}$ as a function of projected radius $\hat{r}$
for three different truncated models, each with $c=8$. 
As the threshold entropy progresses from $\hat{K}_c = 0.1$,
characteristic of clusters, through $\hat{K}_c = 0.3$ to 
$\hat{K}_c = 1.0$, characteristic of groups, the 
$\hat{S}(\hat{r}_\perp)$ profile substantially flattens.     
As long as $\Lambda(T)$ does not change dramatically with 
radius, these emission-measure profiles will be nearly identical 
to the surface-brightness profiles.

For comparision, Figure~\ref{sbright} also shows some $\beta$-model
profiles selected to match the modified-entropy models.  These
$\beta$-model profiles are defined by $\hat{S} \propto [ 1 +
(r/r_c)^2 ]^{-3\beta + 1/2}$, where $r_c$ is the core radius
and $\beta$ determines the asymptotic slope (Cavaliere \& 
Fusco-Femiano 1978).  In each
case, the $\beta$-model closely tracks the modified-entropy 
model within $0.3 r_{200}$.  Beyond that radius, the emission

% ------------------ fig -----------------
\vspace{\baselineskip}
%\begin{figure}
\epsfxsize=3in 
\centerline{\epsfbox{f5.epsi}}
\figcaption{\footnotesize
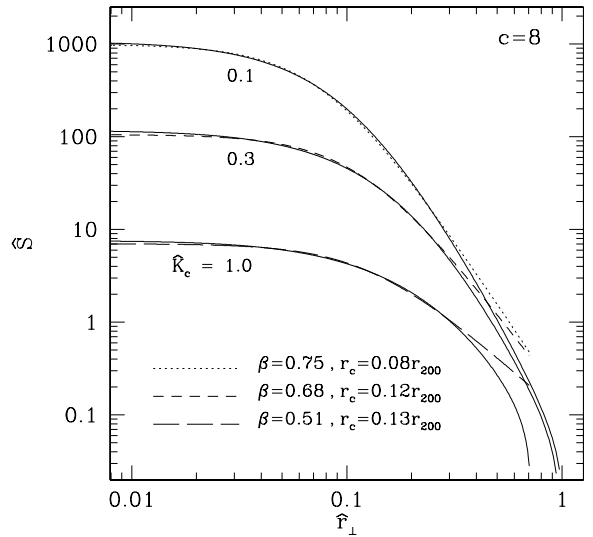
Dimensionless emission measure $\hat{S}$ 
as a function of projected radius $\hat{r}_\perp$.  The solid
lines show $\hat{S}(\hat{r})$ for three modified entropy models
with $c=8$, truncated at $\hat{K}_c = 0.1$, 0.3, and 1.0.  The
dotted and dashed lines show emission measure profiles derived
from $\beta$-models with $\hat{S} \propto [1 + (r/r_c)^2]^{-3\beta
+ 0.5}$.  Notice that the $\beta$-models are very good approximations
to the modified-entropy models within $0.3 r_{200}$.
\label{sbright}}
\vspace{\baselineskip}
%\end{figure}
% ----------------------------------------

\noindent
measures
of the $\beta$-models exceed those of the modified-entropy models,
but owing to the low surface brightness at large radii, the 
$\beta$-model parameters of clusters and groups
have generally been measured within $0.3-0.5 r_{200}$.
Recent measurements by Vikhlinin, Forman, \& Jones (1999)
suggest that cluster surface brightness does indeed fall 
below the best fitting $\beta$-model near the
virial radius.

In order to determine how the best-fitting $\beta$ and $r_c$ depend
on $c$ and $\hat{K}_c$, we performed surface-brightness weighted
fits over a grid of modified entropy models.  Because the best-fit
parameters depend somewhat on the portion of the profile being 
fitted, we elected to do these fits over two different radial
intervals:  $0.03 < \hat{r}_\perp < 0.3$ and $0.05 < \hat{r}_\perp
< 1.0$.  Results for truncated ($\hat{K}_{\rm T}$) models are
shown in Figure~\ref{betafits_t}, and those for shifted 
($\hat{K}_{\rm S}$) models are shown in Figure~\ref{betafits_s}.
Because the profiles in the shifted models tend to be shallower,
the corresponding $\beta$ values are somewhat smaller, particularly
for low values of $\hat{K}_c$.  However, in both cases, the 
best-fitting $\beta$ values span the observed range for clusters
and groups of galaxies (see \S~\ref{sec-sbprofs}).

\subsection{Entropy Modification and Cluster Temperature}
\label{emodtemp}

One consequence of raising the core entropy of a cluster 
is a boost in the luminosity-weighted temperature of the 
intracluster gas.  Figure~\ref{tprofs} shows that increasing
$\hat{K}_c$ raises the temperature of gas at all radii in
both the shifted and truncated models.  Amplification of the
central temperature can be quite extreme, exceeding
$3 T_{200}$ when $\hat{K}_c$ rises to 1.0.  However, the 
resulting enhancement of the luminosity-weighted temperature 
is not nearly as dramatic.  The value of $\hat{T}_{\rm lum}$ 
turns out to depend more critically on the concentration 
parameter $c$ than on the entropy threshold, as demonstrated
in Figure~\ref{tlums_ts}.

% ------------------ fig -----------------
\vspace{\baselineskip}
%\begin{figure}
\epsfxsize=3in 
\centerline{\epsfbox{f6.epsi}}
\figcaption{\footnotesize
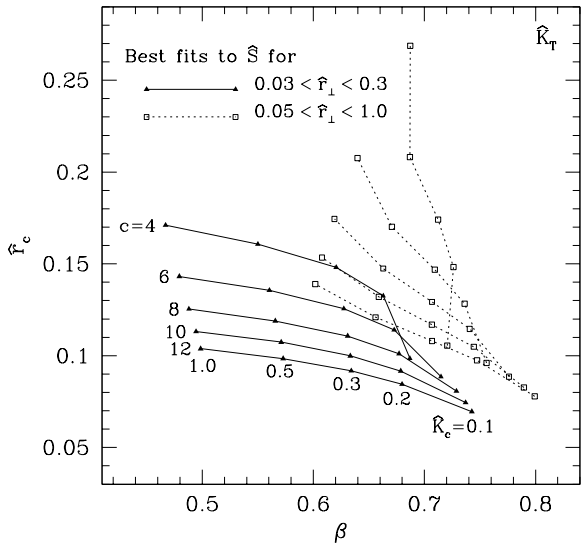
Best-fitting $\beta$-model parameters
for truncated modified-entropy models.  Each line gives the best-fitting
$(\beta,r_c)$ pair for a particular value of the concentration $c$.
The points on each line show best fits for five different values
of the entropy threshold.  The progression $\hat{K}_c = 1.0$, 0.5, 0.3, 
0.2, 0.1 runs from the upper left to lower right in each case.
Because the best fits depend somewhat on the fitted interval, we
depict best fits over the interval $0.03 < \hat{r}_\perp < 0.3$
with solid triangles and solid lines and best fits over the interval 
$0.05 < \hat{r}_\perp < 1.0$ with empty squares and dotted lines.
\label{betafits_t}}
\vspace{\baselineskip}
%\end{figure}
% ----------------------------------------

\vspace*{2em}

% ------------------ fig -----------------
\vspace{\baselineskip}
%\begin{figure}
\epsfxsize=3in 
\centerline{\epsfbox{f7.epsi}}
\figcaption{\footnotesize
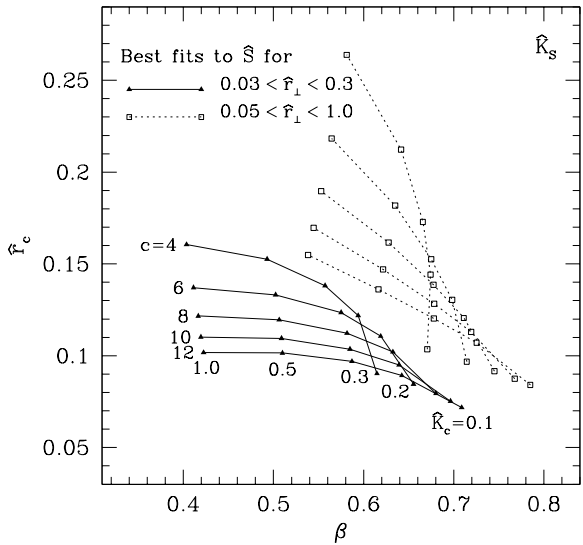
Best-fitting $\beta$-model parameters
for shifted modified-entropy models.  Each line gives the best-fitting
$(\beta,r_c)$ pair for a particular value of the concentration $c$.
The points on each line show best fits for five different values
of the entropy threshold.  The progression $\hat{K}_c = 1.0$, 0.5, 
0.3, 0.2, 0.1 runs from the upper left to lower right in each case.
Because the best fits depend somewhat on the fitted interval, we
depict best fits over the interval $0.03 < \hat{r}_\perp < 0.3$
with solid triangles and solid lines and best fits over the interval 
$0.05 < \hat{r}_\perp < 1.0$ with empty squares and dotted lines.
\label{betafits_s}}
\vspace{\baselineskip}
%\end{figure}
% ----------------------------------------

\vspace*{2em}

The reason that raising the core entropy of gas in an NFW 
potential has only a modest effect on its luminosity-weighted 
temperature involves both the gas density profile

% ------------------ fig -----------------
\vspace{\baselineskip}
%\begin{figure}
\epsfxsize=3in 
\centerline{\epsfbox{f8.epsi}}
\figcaption{\footnotesize
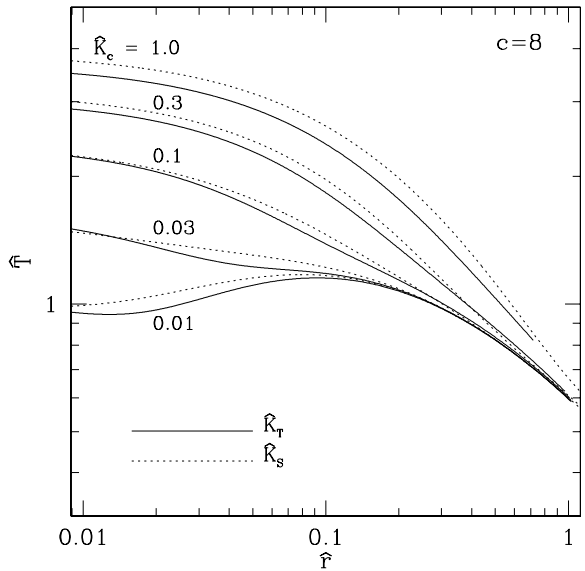
Dimensionless temperature profiles. 
Solid lines show how temperature ($\hat{T}$) depends on
radius ($\hat{r}$) in models with truncated entropy distributions
($\hat{K}_{\rm T})$.  Dotted lines show how temperature depends on
radius in models with shifted entropy distributions
($\hat{K}_{\rm S}$).  Raising the entropy threshold $\hat{K}_c$
increases the temperature at all radii for both kinds of entropy
modification.  This increase is slightly larger for shifted models
because the extra gas at large radii leads to slightly higher
central pressures.
\label{tprofs}}
\vspace{\baselineskip}
%\end{figure}
% ----------------------------------------

\vspace*{2em}

\noindent
and the
dark-matter density profile.  Extra entropy flattens the 
density profile of intracluster gas, diminishing the relative
luminosity of gas in the core.  The characteristic radius
of gas contributing the bulk of the cluster's luminosity
therefore moves outward.  Because the shape of an NFW halo
is steeper than isothermal at $\hat{r} > 1/c$, the characteristic
temperature of gas within such a halo declines outside that radius
(see Figure~\ref{tprofs}).  Thus, the rise in gas temperature
resulting from an increase in $\hat{K}_c$ is largely offset
by an increased contribution to the total luminosity 
from lower-temperature gas at larger radii.  This effect is 
most pronounced in the shifted models, which have substantial 
amounts of relatively cool gas beyond $\hat{r} = 1$.  In those 
models, $\hat{T}_{\rm lum}$ actually {\em declines} as $\hat{K}_c$ 
approaches unity.

The large central temperatures in the shifted and truncated
cases also seem somewhat unrealistic in light of observations
showing either level or radially increasing temperature gradients
within cluster cores (e.g., Arnaud \etal 2001a,b; Allen \etal 2001).  
In order to compare these temperatures
more directly with observations, we must look at $\hat{T}_\perp$,
the luminosity-weighted temperature along various lines of sight
through a cluster.  Figure~\ref{tperp_03} illustrates how
$\hat{T}_\perp$ depends on projected radius $\hat{r}_\perp$
in clusters with $c=8$, $\hat{K}_c = 0.3$, and four different 
kinds of entropy modification.  Projected temperatures at
$\hat{r}_\perp < 0.1$ in both the shifted and truncated models
are still quite high ($\hat{T}_\perp \approx 2 T_{200}$).
However, the core temperatures in the radiative-loss models
look more like those in real clusters (see \S~\ref{sec-tinner}), 
rising with radius in the innermost regions and peaking between 
$1.2 T_{200}$ and $1.6 T_{200}$ at $\sim 0.05 r_{200}$ in 
these particular models.

% ------------------ fig -----------------
\vspace{\baselineskip}
%\begin{figure}
\epsfxsize=3in 
\centerline{\epsfbox{f9.epsi}}
\figcaption{\footnotesize
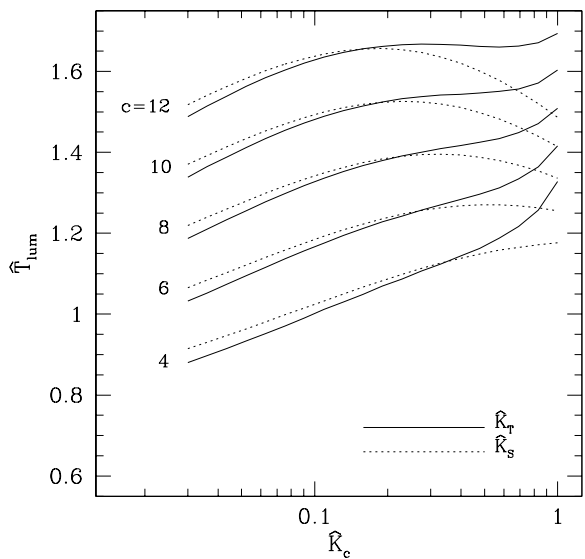
Dimensionless luminosity-weighted temperatures
($\hat{T}_{\rm lum}$) as a function of entropy threshold ($\hat{K}_c$)
for halos of concentration $c = 4$, 6, 8, 10, and 12.  Solid lines
depict truncated models ($\hat{K}_{\rm T}$); dotted lines
depict shifted models ($\hat{K}_{\rm S}$).  The dependence of
$\hat{T}_{\rm lum}$ on $\hat{K}_c$ is rather modest over the
interesting range $0.1 < \hat{K}_c < 1.0$, especially for
larger values of $c$.  This effect arises because a larger 
proportion of the emission from high-$\hat{K}_c$ clusters
comes from $\hat{r} > 1/c$, where temperatures are lower than
in the core.  This outward shift of surface brightness largely
offsets the rise in temperature at all radii evident in 
Figure~\ref{tprofs}.
\label{tlums_ts}}
\vspace{\baselineskip}
%\end{figure}
% ----------------------------------------

\vspace*{1.5cm}

% ------------------ fig -----------------
\vspace{\baselineskip}
%\begin{figure}
\epsfxsize=3in 
\centerline{\epsfbox{f10.epsi}}
\figcaption{\footnotesize
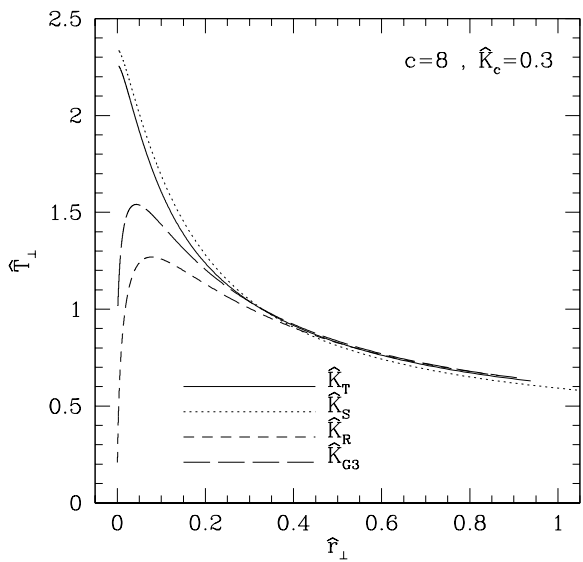
Dimensionless luminosity-weighted line-of-sight
temperature $\hat{T}_\perp$ as a function of projected radius 
$\hat{r}_\perp$.  The solid and dotted lines representing truncated
and shifted models, respectively, both begin at rather high
temperature ($\approx 2 T_{200}$) at small radii and proceed to
decline monotonically with radius.  The long- and short-dashed
lines representing radiative-loss models (see \S~\ref{radloss})
begin at low temperature at $\hat{r} = 0$ and rise to a maximum
at $\approx 0.05 r_{200}$, in better accord with observed 
clusters (see \S~\ref{sec-tinner}).
\label{tperp_03}}
\vspace{\baselineskip}
%\end{figure}
% ----------------------------------------

% ------------------ fig -----------------
\vspace{\baselineskip}
%\begin{figure}
\epsfxsize=3in 
\centerline{\epsfbox{f11.epsi}}
\figcaption{\footnotesize
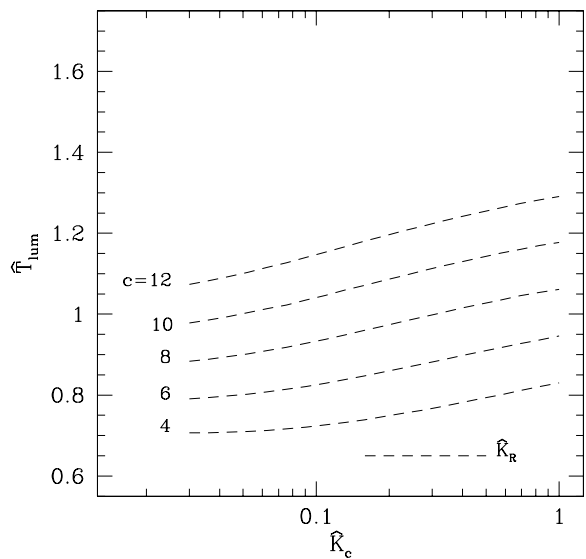
Dimensionless luminosity-weighted temperatures
for the standard radiative-loss model ($\hat{K}_{\rm R}$).  Comparision
of these $\hat{T}_{\rm lum}(\hat{K}_c)$ relations with those in 
Figure~\ref{tlums_ts} shows that emission from the core gas lowers
the luminosity-weighted temperature by $\sim$30\%, relative to
the shifted and truncated models.
\label{tlums_r}}
\vspace{\baselineskip}
%\end{figure}
% ----------------------------------------

\vspace*{2.5cm}

% ------------------ fig -----------------
\vspace{\baselineskip}
%\begin{figure}
\epsfxsize=3in 
\centerline{\epsfbox{f12.epsi}}
\figcaption{\footnotesize
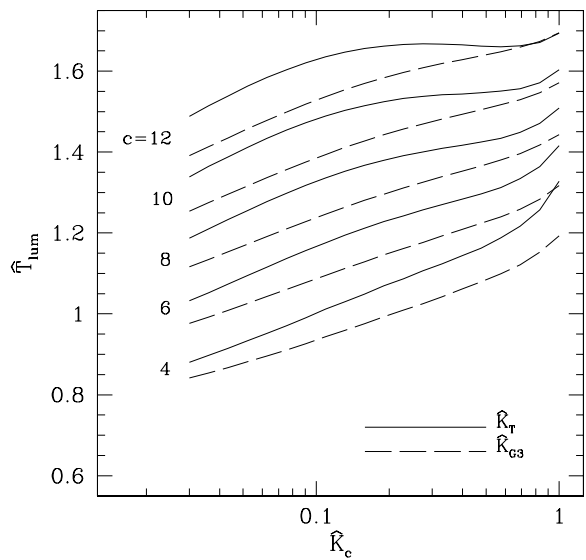
Dimensionless luminosity-weighted temperatures for generalized 
radiative-loss model $\hat{K}_{\rm G3}$ with $\alpha = 3$.  
Solid lines show the $\hat{T}_{\rm lum}(\hat{K}_c)$ relations for
truncated models, and long-dashed lines show the corresponding 
relations for the radiative-loss models.  Luminosity-weighted 
temperature is $\sim$10\% smaller in these generalized radiative-loss
models.  The difference is less dramatic than in the $\alpha = 3/2$ 
case because the amount of gas below $\hat{K}_c$ is smaller,
resulting in somewhat higher core temperature and somewhat lower 
core density and luminosity.
\label{tlums_ta}}
\vspace{\baselineskip}
%\end{figure}
% ----------------------------------------

\clearpage

Cool gas in the cores of clusters with entropy modified through
radiative losses can significantly affect the luminosity-weighted 
temperature if the fraction of gas below $\hat{K}_c$ is relatively 
large.  Figure~\ref{tlums_r} shows that $\hat{T}_{\rm lum}$ for a 
given concentration $c$ and entropy threshold $\hat{K}_c$ is
$\sim$30\% lower in the standard radiative-loss model ($\hat{K}_{\rm R}$)
than in the corresponding truncated model.  This effect is also
present but not nearly so strong in the generalized radiative-loss
model with $\alpha = 3$, where $\hat{T}_{\rm lum}$ is $\sim$10\% 
lower than in the truncated models (see Figure~\ref{tlums_ta}).

\subsection{Entropy Modification and Cluster Luminosity}
\label{emodlum}

All of our schemes for entropy modification substantially
lower the luminosities of clusters.  Figure~\ref{lints_8}
shows how dimensionless luminosity $\hat{L}$ depends on
the entropy threshold $\hat{K}_c$ for clusters with $c=8$.
In all cases, luminosity declines by over an order of magnitude
as $\hat{K}_c$ rises from 0.1 to 1.  This decline is
shallowest for the shifted model because the additional
gas at large radii contributes significantly to the total
luminosity when $\hat{K}_c \sim 1$.  Note also that $\hat{L}$
is over twice as large in the standard radiative-loss
case ($\hat{K}_{\rm R}$) because a large proportion of
the luminosity in such models comes from the dense core.
Luminosity enhancement in the $\alpha=3$ radiative-loss model
is more moderate---only a few tens of percent greater than 
in the truncated case.

Analyzing how $\hat{L}$ depends on $\hat{K}_c$ leads to some 
important insights into how the cooling threshold determines 
the luminosity-temperature relation of clusters and groups.
Figure~\ref{lints_tx} shows that $\hat{L} \propto \hat{K}_c^{-3/2}$
is a good approximation to the $\hat{L}$-$\hat{K}_c$ relation
for truncated models in the range $0.1 < \hat{K}_c < 1.0$.
This scaling reflects the asymptotic slope of the NFW
density profile.  Notice that the outer portions of the density 
profiles in Figure~\ref{denprofs} track the underlying
NFW profile of the dark matter, and the inner portions are
nearly isentropic with $\hat{K} \approx \hat{K}_c$.  One can 
therefore approximate the density profile of a truncated
model by joining an isentropic density profile to an NFW
profile at the radius $\hat{r}_K$ at which the density 
of the unmodified NFW profile is $\hat{\rho}_K \sim 
\hat{K}_c^{-3/2}$.  The analytical isentropic solution
from \S~\ref{dimensionless} tells us that $\hat{\rho}
\stackrel {\propto} {\sim} [\hat{r}^{-1} \ln (1+c\hat{r})]^{3/2}$ 
within $\hat{r}_K$, so the luminosity from the isentropic 
gas scales $\propto \hat{\rho}_K^2 \hat{r}_K^3$.  The luminosity 
from the NFW portion of the profile scales in the same
way for the radii of interest.  Thus, for larger values
of $\hat{r}$, where the density of the NFW profile
approaches $\hat{\rho} \propto \hat{r}^{-3}$, we have
$\hat{L} \propto \hat{\rho}_K \propto \hat{K}_c^{-3/2}$.

This particular scaling of $\hat{L}$ with $\hat{K}_c$ has some
very interesting consequences.  Section~\ref{integrated}
showed that the physical luminosity of a cluster scales
as $L \propto r_{200}^3 \rho_{cr}^2 \Lambda \hat{L}$.
In order to substitute physical quantities for $\hat{L}$,
we need to recognize that $\hat{K}_c^{3/2} \propto K_c^{3/2}
T_{200}^{-3/2} \rho_{cr}$ and that $K_c^{3/2} \propto T^{1/2}
\Lambda t_{\rm H}$, where $t_{\rm H}$ is the age of the universe.
We therefore obtain
\begin{equation}
\label{ltscal}
 L \, \propto \, T_{\rm lum}^{5/2} \, \hat{T}_{\rm lum}^{-3} 
              \, (Ht_{\rm H})^{-1} \; \; ,
\end{equation}
where $H$ is the Hubble constant at time $t_{\rm H}$.

% ------------------ fig -----------------
\vspace{\baselineskip}
%\begin{figure}
\epsfxsize=3in 
\centerline{\epsfbox{f13.epsi}}
\figcaption{\footnotesize
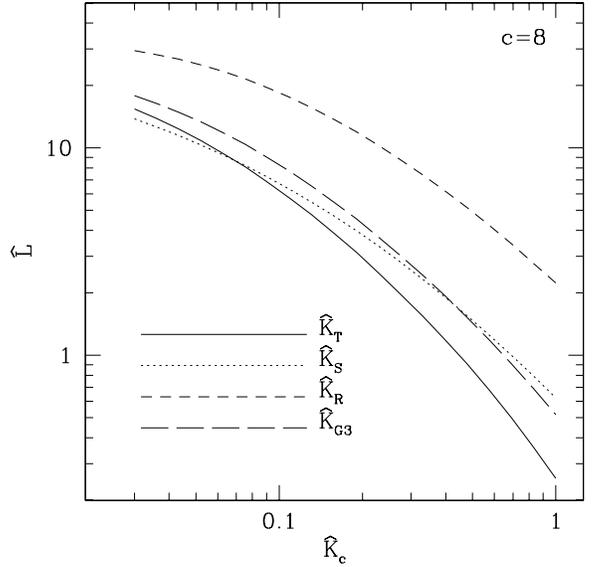
Dependence of dimensionless luminosity $\hat{L}$
on the entropy threshold $\hat{K}_c$ for clusters with
$c=8$.  The decline of $\hat{L}$ with increasing $\hat{K}_c$
is strong for all types of entropy modification.  The effect
is stronger in the truncated ($\hat{K}_{\rm T}$) and
radiative-loss ($\hat{K}_{\rm R}$, $\hat{K}_{\rm G3}$) 
models than in the shifted model ($\hat{K}_{\rm S}$)
because the latter has more gas at large radii, which
contributes significantly to $\hat{L}$ when the entropy
threshold is high.
\label{lints_8}}
\vspace{\baselineskip}
%\end{figure}
% ----------------------------------------

\vspace*{1.5cm}

% ------------------ fig -----------------
\vspace{\baselineskip}
%\begin{figure}
\epsfxsize=3in 
\centerline{\epsfbox{f14.epsi}}
\figcaption{\footnotesize
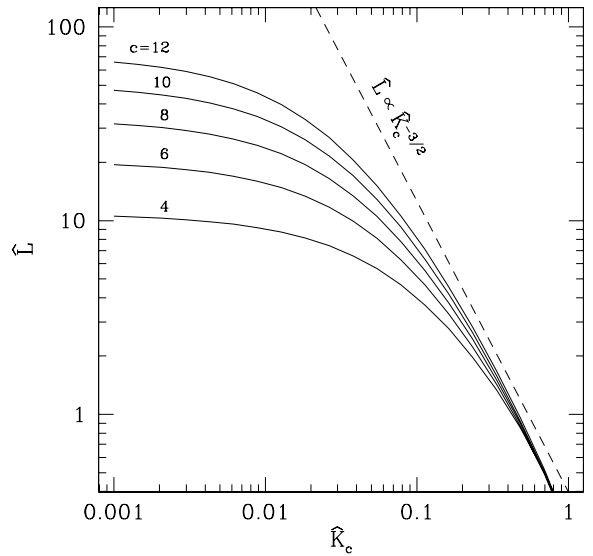
Dimensionless luminosity $\hat{L}$ as a function of entropy
threshold $\hat{K}_c$ in truncated ($\hat{K}_{\rm T}$) models
for clusters with concentration $c=4$, 6, 8, 10, and 12.  
The cluster luminosities in these models
sharply decline as $\hat{K}_c$ increases.  For $0.1 < \hat{K}_c
< 1.0$ this decline approximately scales as $\hat{L} \propto
\hat{K}_c^{-3/2}$.
\label{lints_tx}}
\vspace{\baselineskip}
%\end{figure}
% ----------------------------------------

\clearpage

The following three features of this expression are particularly
noteworthy:
\begin{itemize}
\item The power-law slope it predicts for the X-ray 
      $L \propto T^b$ relation is very close to the 
      observed slope of $b \approx 2.6-2.9$ 
      (e.g., Markevitch 1998; Arnaud \& Evrard 1999).  
      Because the halos of low-temperature groups have 
      higher concentrations than high-temperature clusters
      (see \S~\ref{conparam}), we expect $\hat{T}_{\rm lum}$ 
      to decline by a factor $\sim 1.3$ as $T_{\rm lum}$
      rises by an order of magnitude.  This dependence
      on concentration steepens the power-law slope
      implied by equation (\ref{ltscal}) to 
      $b \approx 2.8$.
\item The $L$-$T$ relation does not depend on the form 
      of the cooling function, which scales as $\Lambda
      \propto T^{1/2}$ in the free-free cooling regime
      at $T > 2 \, \keV$ and as $\Lambda \propto T^{-1/2}$ 
      in the line-cooling regime at $T < 2 \, \keV$.
      If the entropy threshold in real clusters did 
      not depend on $\Lambda$, then we would expect a
      distinct steepening of the $L$-$T$ relation as $T$
      rises through $\sim 2 \, \keV$, which is not
      observed.  Equation (\ref{ltscal}) is indepedent
      of $\Lambda$ because the dimensionless gas mass
      within $\hat{r}_K$ scales $\propto \hat{\rho}_K 
      \hat{r}_K^3$, and thus remains approximately constant
      in an NFW halo for $\hat{r}_i > 1/c$.  The cooling 
      time of that gas is $\sim t_{\rm H}$, so its physical
      luminosity scales $\stackrel {\propto} {\sim} 
      M_{200} T_{200} t_{\rm H}^{-1}$.  This scaling law
      also applies to the entire cluster, because the
      luminosity of gas outside $\hat{r}_K$ scales in
      lockstep with gas inside $\hat{r}_K$.  Applying the
      relation $M_{200} \propto T_{200}^{3/2} H^{-1}$ 
      thus leads back to equation (\ref{ltscal}), except
      for the $\hat{T}_{\rm lum}$ factor.
\item Finally, the $L$-$T$ relation implied by equation
      (\ref{ltscal}) changes very little with time, again
      in agreement with observations (Mushotzky \& Scharf 1997;
      Donahue \etal 1999; Della Ceca \etal 2000; 
      Borgani \etal 2001b).  That happens
      because the mass of a cluster of a given temperature
      declines $\propto H^{-1}$ with increasing redshift. 
      According to the 
      $L \propto M_{200} T_{200} t_{\rm H}^{-1}$ scaling,
      the resulting decline in the intracluster gas mass
      compensates almost precisely for the drop in the
      cooling threshold owing to the shorter cooling 
      time $t_{\rm H}$.
\end{itemize}

This scaling breaks down at the low-temperature end,
where $\hat{K}_c \gtrsim 1$, because the entire intracluster
(or intragroup) medium in the truncated and shifted cases
becomes effectively isentropic.  Then, the gas density determined 
by the cooling threshold is $\propto T / \Lambda t_{\rm H}$, 
and the luminosity from within $r_{200}$ scales as $L \propto 
T_{\rm lum}^{7/2} \Lambda^{-2} \hat{T}_{\rm lum}^{-3/2} (Ht_{\rm H})^{-2} 
H^{-1}$.  Thus, we should expect the $L$-$T$ relation for 
low-temperature halos to be steeper than for clusters, roughly 
$L \propto T^{4.5}$ for $\Lambda \stackrel {\propto} {\sim} 
T^{-1/2}$, and we should expect the luminosity of objects 
at a given temperature to increase with time.

In order to make further progress, we will need to understand
how halo concentration depends on halo mass.

\section{Modeling Real Clusters}

The previous section explored the properties of dimensionless
modified-entropy models for clusters depending on two parameters, 
the halo concentration parameter $c$ and an entropy-threshold 
parameter $\hat{K}_c$.  As in Voit \& Bryan (2001), we argued that 
the crucial entropy threshold should be the entropy level at 
which intracluster gas would cool in a Hubble time.  Now we will 
use a set of relations between halo concentration and halo mass 
to construct modified-entropy models for real clusters.  In these 
models, the halo mass $M_{200}$ determines the concentration 
parameter $c$.  The mass and halo concentration then set the 
temperature of the cluster, and this temperature sets the level 
of the entropy threshold $K_c$.  Cluster properties are therefore 
determined primarily by the overall cosmological model, because 
it fixes the global baryon fraction, the relation between halo 
mass and concentration, and the Hubble time that
governs $K_c(T)$.  Some of a cluster's properties, such as
the temperature of the innermost gas, also depend on the chosen 
scheme for entropy modification.  The properties of the resulting 
modified-entropy models turn out to be remarkably similar to 
those of observed clusters and groups.

The section begins by defining the $M_{200}$-$c$ relations we will 
use and comparing them to some of the scarce data.  Then, we explore 
the structure of the intracluster medium in our modified-entropy 
cluster models, showing that the observed relationship
between $\beta$ and $T_{\rm lum}$ may be more than just a selection 
effect and demonstrating how entropy modification can lead to 
temperature profiles similar to those observed in both the inner 
and outer regions of clusters.
We go on to compare the mass-temperature relation from our 
modified-entropy models to observations, showing that the models 
reproduce the observational results of Nevalainen \etal (2000) 
and Finoguenov \etal (2001).  We also compare the $L_X$-$T_{\rm lum}$
relation from our models to cluster data, again showing excellent
agreement.  Much of the dispersion of this relation, in both the
models and the data, stems from differences in the amount of gas
below the cooling threshold, and we conclude with some speculations
about what governs these differences.

\subsection{Concentration Parameter}
\label{conparam}

The concentration of a dark matter halo is
determined by its formation history.  Early formation 
generally leads to a denser core and thus a higher 
concentration.  Low-mass halos, which tend to collapse 
earlier in time, are therefore expected to be somewhat more 
concentrated than high-mass halos.  Because a halo's collapse
and merger history determines its concentration, the relation 
between concentration and halo mass depends on the 
underlying cosmological model, specified by the matter 
density $\Omega_M$, the dark-energy density $\Omega_\Lambda$, 
the normalization $\sigma_8$ of the perturbation spectrum, 
and the shape of that spectrum, often parameterized by 
$\Gamma$ (Bardeen \etal 1986; Sugiyama 1995).  
In this paper, we will restrict our 
attention to currently-favored $\Lambda$CDM models, in 
which the resulting clusters turn out to be 
thoroughly consistent with observations.

Several versions of the concentration parameter $c_\Delta$ can
be found in the literature, each defined with respect to
a radius $r_\Delta$ within which the mean halo density
is $\Delta \rho_{cr}$.  Here we will define $c$ with
respect to $r_{200}$, in order to be consistent with the
models developed in the previous section; thus, $c \equiv c_{200}$.
Figure~\ref{c200} shows various relations between halo mass and
$c_{200}$ that can be found in the lit-\clearpage

% ------------------ fig -----------------
\vspace{\baselineskip}
%\begin{figure}
\epsfxsize=3in 
\centerline{\epsfbox{f15.epsi}}
\figcaption{\footnotesize
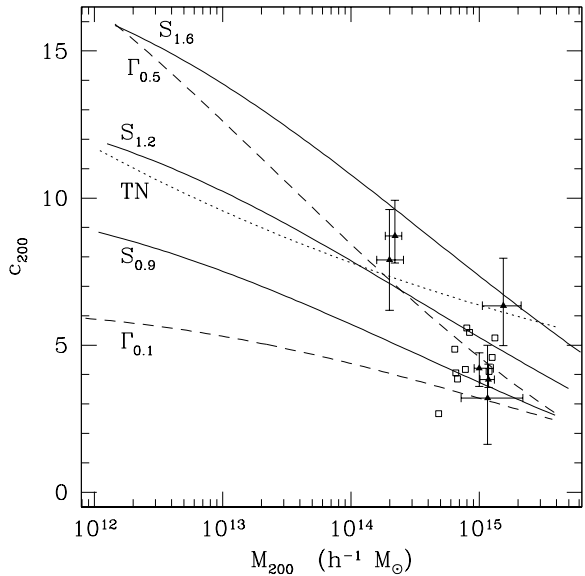
Concentration parameter ($c_{200}$) as a function of 
halo mass ($M_{200}$).  Solid triangles give concentrations
derived from {\em Chandra} observations of real clusters
by Allen \etal (2001).  Empty squares give the concentrations
of clusters simulated by Eke, Navarro, \& Frenk (1998) in a $\Lambda$CDM
cosmology with $\sigma_8 = 1.1$.  Solid lines show $c_{200}$
values predicted by models $S_{0.9}$, $S_{1.2}$, and $S_{1.6}$
from Eke \etal (2001), for a $\Lambda$CDM cosmology in which 
$\sigma_8 = 0.9$, 1.2, and 1.6, respectively, for a power spectrum
with shape parameter $\Gamma = 0.2$.  Dashed lines show $c_{200}$
given by models $\Gamma_{0.1}$ and $\Gamma_{0.5}$ from Eke \etal 
(2001), for which $\Gamma = 0.1$ and 0.5, respectively, and 
$\sigma_8 = 0.9$.  The dotted line labeled TN shows the $c_{200}$ 
values from the fit of Tozzi \& Norman (2001) to the NFW 
prescription for a $\Lambda$CDM cosmology with $\sigma_8 = 1.1$.
\label{c200}}
\vspace{\baselineskip}
%\end{figure}
% ----------------------------------------

\noindent
erature.
The {\em Chandra} observations of Allen \etal (2001) suggest 
that concentrations corresponding to $c_{200} \approx 4-6$ are 
appropriate for massive ($\sim 10^{15} \, M_\odot$) clusters. 
Concentrations of groups are less certain because they
are more model-dependent, and the dispersion in $c_{200}$ is
expected to increase at lower halo masses (Afshordi \& Cen 2002).

In this section, we will adopt the halo mass-concentration
relations predicted by models $S_{0.9}$ and $S_{1.2}$ from 
Eke, Navarro, \& Steinmetz (2001), which are consistent 
with the {\em Chandra}
data on massive clusters.  Both models are derived from 
$N$-body simulations of $\Lambda$CDM cosmologies
with $\Omega_M = 0.3$, $\Omega_\Lambda = 0.7$, and
$h = 0.65$.  The power spectrum in each model has
a shape parameter $\Gamma = 0.2$, but the normalization
of the power spectrum differs:  $\sigma_8 = 0.9$ in model 
$S_{0.9}$ and $\sigma_8 = 1.2$ in model $S_{1.2}$.  The higher 
normalization in model $S_{1.2}$ leads to halos that are 
about 30\% more concentrated.  Figure~\ref{kc_m200} shows
the dependence of $\hat{K}_c$ on halo mass that follows from
model $S_{1.2}$; the entropy thresholds for model $S_{0.9}$ 
are very similar.

\subsection{Cluster Structure}

The cluster models we analyze in this section are completely 
determined by the assumed relationship between halo mass and 
concentration (either $S_{0.9}$ or $S_{1.2}$) and the chosen
scheme for entropy modification ($\hat{K}_T$, $\hat{K}_S$, 
$\hat{K}_R$, or $\hat{K}_{G3}$).  For each halo mass, we 
compute the bolometric X-ray luminosity $L_X$, luminosity-weighted
temperature $T_{\rm lum}$, X-ray surface-brightness profile 
$S_X(r_\perp)$, and projected temperature profile 
$T_\perp(r_\perp)$.  Unlike in \S~\ref{integrated}, each integral
includes a cooling function $\Lambda(T)$ drawn from the

% ------------------ fig -----------------
\vspace{\baselineskip}
%\begin{figure}
\epsfxsize=3in 
\centerline{\epsfbox{f16.epsi}}
\figcaption{\footnotesize
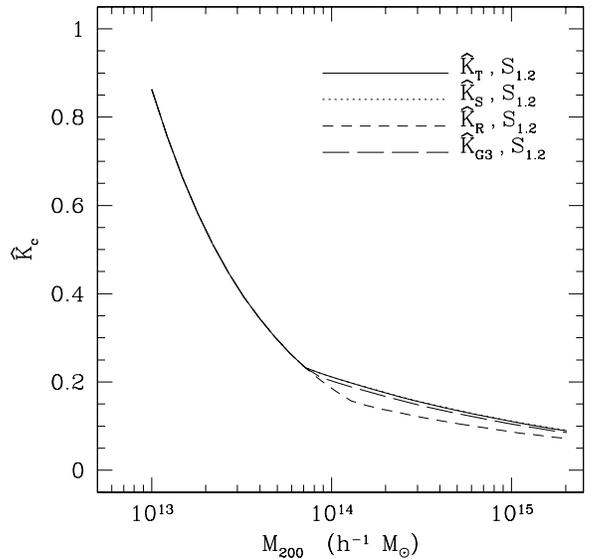
Dimensionless entropy threshold ($\hat{K}_c$) determined by
cooling as a function 
of halo mass $M_{200}$.  The lines show the entropy thresholds
for truncated models (solid line), shifted models (dotted line),
and radiative-loss models with $\alpha = 3/2$ (short-dashed line),
and $\alpha = 3$ (long-dashed line), assuming the 
halo mass-concentration relation from model $S_{1.2}$.  The
corresponding thresholds for model $S_{0.9}$ are very similar.
Above $10^{14} \, h^{-1} \, M_\odot$, different entropy-modification
schemes produce slightly different dimensionless thresholds because
the temperature $T_{\rm lum}$ associated with a given halo mass is
slightly different.  Equations~(\ref{eq-Kc}) and (\ref{eq-hatKc})
give the entropy threshold $K_c$ and dimensionless entropy threshold
$\hat{K}_c$, respectively, as functions of temperature.
\label{kc_m200}}
\vspace{\baselineskip}
%\end{figure}
% ----------------------------------------

\noindent
models of Raymond \& Smith (1977) for
gas with 30\% solar metallicity.  We will first 
focus on the spatially resolved
properties of these
model clusters, showing how their surface-brightness profiles
and temperature gradients compare with those 
of observed clusters.

\subsubsection{Surface-Brightness Profiles}
\label{sec-sbprofs}

Modified-entropy models of groups tend to have flatter
surface-brightness profiles than hot clusters, in accordance
with observations (e.g., Ponman \etal 1999; Horner \etal 1999).
Because the surface-brightness profiles of both
real clusters and our modified-entropy models are well
described by $\beta$-models with $S_X \propto [1 + 
(r/r_c)^2]^{-3\beta + 1/2}$ inside of $0.3 \, r_{200}$
(see \S~\ref{emodsb}), we compare our models with the data 
by comparing the best-fitting $\beta$-model parameters.
Figure~\ref{bmods} shows how $\beta$ and $r_c$
depend on $T_{\rm lum}$ in truncated modified-entropy
models with relation $S_{1.2}$ and in the cluster
data of Finoguenov \etal (2001).  The models reproduce
the observed range of both parameters and the observed
rise in each parameter with increasing temperature.
However, the observed dispersion in each parameter
at a given temperature is quite large.  Wu \& Xue (2002) 
find the same kind of temperature dependence in their
cluster models, based on the work of Bryan (2000), which 
are very similar to the truncated models developed 
in this paper.

Some of the dispersion in $\beta$ and $r_c$ at a given temperature
may stem from selection effects.
Because the surface-brightness profiles of modified-entropy clusters
continually steepen with radius, unlike those of $\beta$-models,
the best-fitting $\beta$ and $r_c$ depend somewhat on the range
of radii being fit.  Figure~\ref{bmods} shows that the best-fitting
values of $\beta$ and $r_c$ both decline as the surface-brightness
threshold of  \clearpage

% ------------------ fig -----------------
\vspace{\baselineskip}
%\begin{figure}
\epsfxsize=3in 
\centerline{\epsfbox{f17.epsi}}
\figcaption{\footnotesize
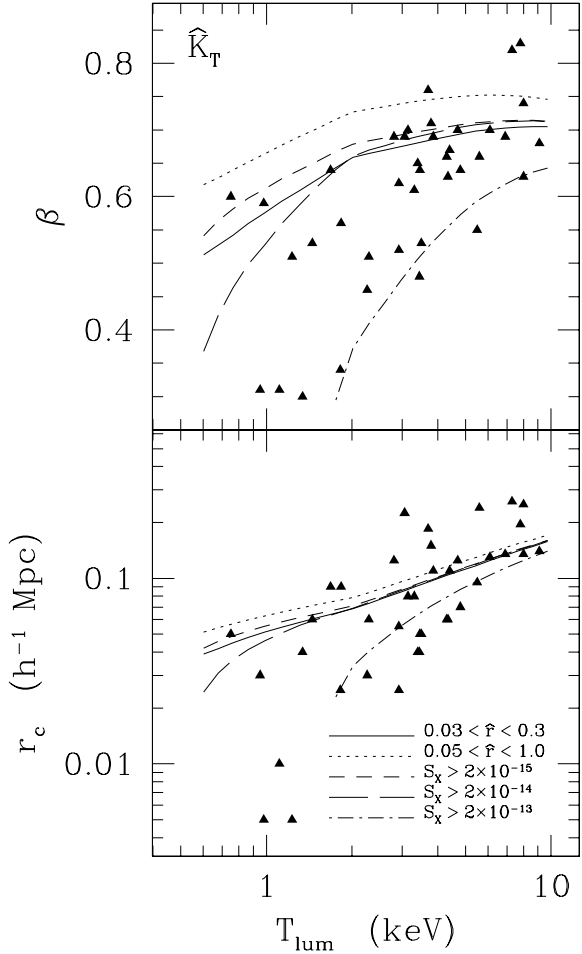
Best-fitting $\beta$-model parameters as a function of 
luminosity-weighted temperature $T_{\rm lum}$.  In a $\beta$-model, 
X-ray surface brightness declines with radius as $S_X \propto 
[1+(r/r_c)^2]^{-3\beta+1/2}$, where $r_c$ is the core radius and 
$\beta$ determines the asymptotic slope.  Triangles show 
measured values of $\beta$ (upper panel) and $r_c$
(lower panel) compiled by Finoguenov \etal (2001).  
Solid lines show the values of these parameters in the
best fits to truncated modified entropy models ($\hat{K}_T$)
over the radial range $0.03<\hat{r}<0.3$, where $\hat{r} \equiv
r/r_{200}$.  Dotted lines show the best-fitting values over
the range $0.05<\hat{r}<1.0$.  The other lines show best fits
above surface-brightness thresholds of $2 \times 10^{-15} \, 
{\rm erg \, cm^{-2} \, s^{-1} \, arcmin^{-2}}$ (short-dashed 
lines), $2 \times 10^{-14} \, {\rm erg \, cm^{-2} \, s^{-1} \, 
arcmin^{-2}}$ (long-dashed lines), and $2 \times 10^{-13}
\, {\rm erg \, cm^{-2} \, s^{-1} \, arcmin^{-2}}$ (dot-dashed
lines).
\label{bmods}}
\vspace{\baselineskip}
%\end{figure}
% ----------------------------------------

\vspace*{1em}

\noindent
the fit rises from $2 \times 10^{-15} \, \sbm$ 
through $2 \times 10^{-14} \, \sbm$ to $2 \times 10^{-13} \, {\rm erg \,
cm^{-2} \, s^{-1}}$ ${\rm arcmin^{-2}}$.  This 
effect is more severe for groups than for clusters because their
overall surface brightness is smaller.  The figure also shows that
fits over the radial range $0.05 < r/r_{200} < 1.0$ yield slightly
larger $\beta$ and $r_c$ than fits over $0.03 < r/r_{200} < 0.3$.
Komatsu \& Seljak (2001) noted a similar surface-brightness
selection effect in 
their polytropic models for clusters.  However, the rise of $\beta$ 
with $T_{\rm lum}$ in their models is entirely due to this 
surface-brightness bias.  In our models, it is largely a 
consequence of the greater impact of entropy modification
on groups.  Puzzlingly, this selection effect becomes stronger
at low surface brightness levels in the models of Wu \& Xue (2002),
perhaps because their boundary condition leads to a less precipitous
decline in $S_X$ near the virial radius. 

% ------------------ fig -----------------
\vspace{\baselineskip}
%\begin{figure}
\epsfxsize=3in 
\centerline{\epsfbox{f18.epsi}}
\figcaption{\footnotesize
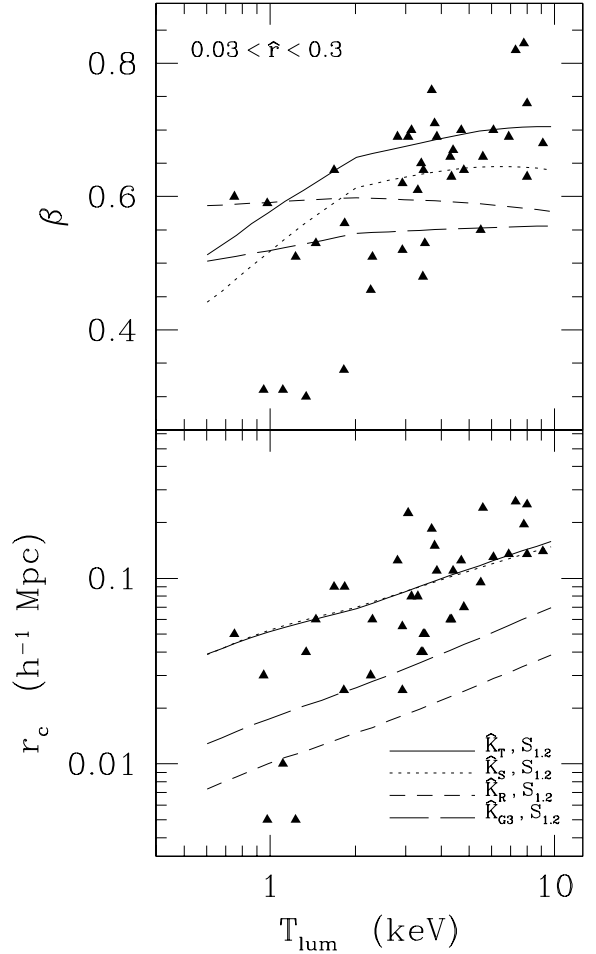
Best-fitting $\beta$-model parameters to various modified entropy
models as a function of luminosity-weighted temperature $T_{\rm lum}$.
As in Figure~\ref{bmods}, triangles show measured values of $\beta$ 
(upper panel) and $r_c$ (lower panel) compiled by Finoguenov \etal (2001).  
Solid lines show the values of these parameters in the
best fits to truncated modified-entropy models ($\hat{K}_T$),
dotted lines show the best-fitting values for shifted models ($\hat{K}_S$),
short-dashed lines show best fits for radiative-loss model $\hat{K}_R$,
and long-dashed lines show best fits for radiative-loss model $\hat{K}_{G3}$.
All fits were performed over the radial range $0.03<\hat{r}<0.3$ for
models based on concentration-parameter relation $S_{1.2}$.
\label{bmods_all}}
\vspace{\baselineskip}
%\end{figure}
% ----------------------------------------

\vspace*{3em}

The observed dispersion in $\beta$ and $r_c$ may also depend
in part on the amount of gas below the cooling threshold.
Figure~\ref{bmods_all} illustrates how $\beta$ and $r_c$
depend on the scheme for entropy modification.  Shifted
models have core radii identical to those of the corresponding
truncated models and slightly smaller values of $\beta$.  However, allowing
for radiative losses reduces the best-fitting core radius by
a factor of $3-5$ and flattens the $\beta$-$T_{\rm lum}$ relation, yielding
$\beta \approx 0.5-0.6$ for fits over the interval $0.03 < r/r_{200}
< 0.3$.  Other possible sources of dispersion include ellipticity
and cluster substructure, which cannot be investigated in the context
of our spherically symmetric models.

% ------------------ fig -----------------
\vspace{\baselineskip}
%\begin{figure}
\epsfxsize=3in 
\centerline{\epsfbox{f19.epsi}}
\figcaption{\footnotesize
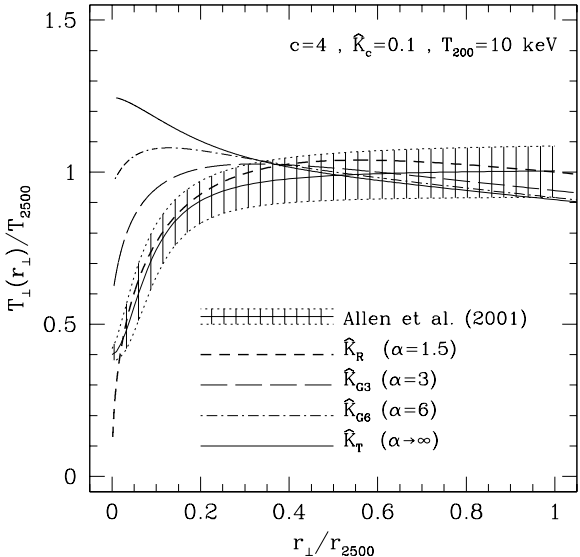
Luminosity-weighted projected temperature $T_\perp$ as a function
of projected radius $r_\perp$.  Radii are given in units of the
radius $r_{2500}$ within which the mean matter density is 2500
times the critical density.  Temperatures are given in units
of the mean mass-weighted temperature $T_{2500}$ within $r_{2500}$.
The hatched area shows the best fit of Allen \etal (2001) to their
Chandra data on six hot clusters with strong cooling flows, 
along with the uncertainty in that fit.
Other lines show $T_\perp$ from modified-entropy models with
parameters characteristic of hot clusters: $c=4$, $\hat{K}_c = 0.1$,
and $T_{200} = 10 \, \keV$.  The short-dashed line depicts
a standard radiative-loss model ($\hat{K}_R$), the long-dashed
line depicts generalized radiative-loss model $\hat{K}_{G3}$ with
$\alpha = 3$, the dot-dashed line shows a generalized
radiative-loss model with $\alpha = 6$, and the solid
line shows a truncated model ($\hat{K}_T$), in which $\alpha$
is effectively infinite.
\label{tinner}}
\vspace{\baselineskip}
%\end{figure}
% ----------------------------------------

\subsubsection{Temperature Gradient at Small Radii}
\label{sec-tinner}

The temperature gradient within $\sim 0.3 \, r_{200}$ also appears
to depend on the amount of gas below the cooling threshold.
Figure~\ref{tinner}, in which the model parameters are appropriate
for hot clusters, shows that $T_\perp(r_\perp)$ gradually declines
with radius when the entropy distribution is simply truncated
at the entropy threshold.  Our radiative-loss models, in contrast,
have temperatures that rise outward from $r=0$, then flatten,
becoming virtually isothermal out to 30\% of the virial radius.
As the parameter $\alpha$ drops from the very large values characteristic
of truncated models to the canonical radiative-loss value of 3/2, 
the region with a positive temperature gradient becomes progressively larger.
The run of temperature with radius in our model with $\alpha = 3/2$
is very similar to the ``universal'' temperature profile observed
by Allen \etal (2001) in clusters that appear to have strong cooling 
flows.  In order to facilitate comparison of our models with that
data, Figure~\ref{tinner} shows radii in units of $r_{2500}$, the radius
within which the mean matter density is $2500 \rho_{cr}$, and temperature
in units of $T_{2500}$, the mean mass-weighted temperature of intracluster
gas within $r_{2500}$.  For clusters of this concentration, $r_{2500}
\approx 0.3 r_{200}$.  Note that setting $\alpha = 6$ produces
a cluster model with a temperature gradient remaining within 10\% 
of isothermal from small radii to beyond $0.3 r_{200}$.

% ------------------ fig -----------------
\vspace{\baselineskip}
%\begin{figure}
\epsfxsize=3in 
\centerline{\epsfbox{f20.epsi}}
\figcaption{\footnotesize
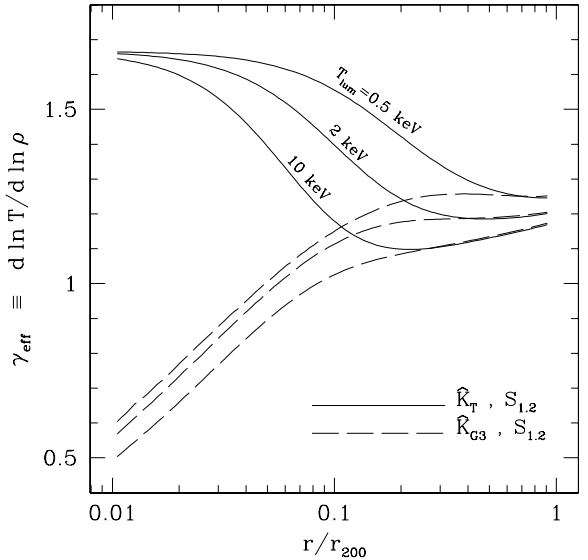
Effective polytropic index $\gamma_{\rm eff} \equiv d \ln P / d \ln \rho$
as a function of radius in modified-entropy models.  Solid lines
depict truncated models ($\hat{K}_T$), and dashed lines depict
radiative-loss models ($\hat{K}_{G3}$) with $\alpha = 3$.
Results are given for models of clusters with three different
luminosity-weighted temperatures ($T_{\rm lum} = 0.5$, 2.0, and $10.0 \,
\keV$, top-to-bottom), using concentration-parameter relation
$S_{1.2}$.  The curves show that the truncated models are isentropic
($\gamma_{\rm eff} = 5/3$) at small radii and nearly isothermal 
($\gamma_{\rm eff} \approx 1.1-1.2$) at large radii.  Values of 
$\gamma_{\rm eff}$ in the radiative-loss models are similar at 
large radii but drop below unity in the radiative-loss models 
because of the low central temperatures.
\label{gamma_profs}}
\vspace{\baselineskip}
%\end{figure}
% ----------------------------------------

\subsubsection{Temperature Gradient at Large Radii}

Projected cluster temperatures at large radii gradually decline
with radius in all of our modified-entropy models.  
{\em ASCA} observations of many clusters show evidence
for such a decline, which is often expressed as an
effective polytropic index $\gamma_{\rm eff} 
\equiv d \ln P / d \ln \rho$ (e.g., Markevitch \etal 1998;
Markevitch \etal 1999; Ettori \& Fabian 1999).  These observations 
indicate that $\gamma_{\rm eff} \approx 1.2$ outside
the cores of clusters.  However, recent {\em XMM-Newton}
observations have shown that at least some clusters are
nearly isothermal within $0.4 \, r_{200}$, except for
the low-temperature core (Arnaud \etal 2001a,b).  The 
results of \S~\ref{sec-tinner} suggest that radiative 
losses may be responsible for the isothermal profiles
found in these clusters.

Figure~\ref{gamma_profs} shows how $\gamma_{\rm eff}$
varies with radius in truncated modified-entropy models
and in radiative-loss model $\hat{K}_{G3}$.  The cores
of the truncated models are nearly isentropic, with
$\gamma_{\rm eff} \approx 5/3$ at $r \lesssim 0.01$.
This isentropic region extends to larger radii in cooler 
halos because their entropy thresholds amount to a larger
fraction of the halo's mean entropy.  Outside the 
isentropic region, we find $\gamma_{\rm eff} \approx
1.1-1.2$, in agreement with observations.  The origin
of this value of $\gamma_{\rm eff}$ can be traced back
to our assumption that the unmodified gas and dark-matter 
density profiles are described by identical NFW laws.  Komatsu 
\& Seljak (2001) have exploited that same assumption to construct 
polytropic models for the intracluster
medium, finding that $\gamma_{\rm eff} \approx 1.1-1.2$
produces the best agreement between these density profiles
at large radii.

The behavior of $\gamma_{\rm eff}$ in the radiative-loss models
is similar to the truncated models at large radii but quite different
at small radii.  Figure~\ref{gamma_profs} shows that the effective
polytropic index steadily rises from $\gamma_{\rm eff} < 1$ inside
the core through $\gamma_{\rm eff} \approx 1$ near the core radius
to $\gamma_{\rm eff} \approx 1.1-1.2$ at larger radii.  The point
at which a cluster makes this transition from an isothermal
gradient to a mildly declining gradient depends somewhat on
cluster temperature.  Hot clusters with massive halos tend to
have smaller concentrations which allow for near-isothermality 
at larger radii.

In order to compare our models with observations, we have evaluated
$\gamma_{\rm eff}$ at $r/r_{200} = 0.2$, essentially twice the
core radius.  This radius is large enough to escape the severe
model-dependence of $\gamma_{\rm eff}$ within the cluster core but
small enough that the signal-to-noise in cluster observations still
allows for accurate measurements.  Figure~\ref{gammas} compares 
those evaluations of $\gamma_{\rm eff}$ with the data of
Finoguenov \etal (2001).  Our models agree with the data,
within the large observational uncertainties, and also 
share the observed (2.5$\sigma$) tendency for $\gamma_{\rm eff}$ 
to decline with increasing temperature.  Two different
effects combine to produce higher values of $\gamma_{\rm eff}$ at 
low temperature.  Because the halos of groups tend to
be more concentrated than those of clusters, the scaled
temperature $T/T_{200}$ tends to be larger in the cores
of groups, leading to higher $\gamma_{\rm eff}$.  An
additional enhancement of $\gamma_{\rm eff}$ appears
in truncated models, in which the entropy threshold
has a more significant effect at $0.2 \, r_{200}$ on
group scales (see Figure~\ref{gamma_profs}).

\subsection{Mass-Temperature Relation}

Many efforts to measure cosmological parameters such as
$\Omega_M$ and $\sigma_8$ have relied heavily on the cluster
mass-temperature relation to link cluster masses to observables
like cluster temperature and luminosity (e.g., Henry 1997; Bahcall
\& Fan 1998; Eke \etal 1998; Donahue \& Voit 1999; Borgani \etal 1999;
Henry 2000).  Until very
recently, observers had to rely on $M$-$T$ relations that
were calibrated with numerical simulations (e.g., Evrard \etal 
1996; Bryan \& Norman 1998).  These simulations provide the
constant of proportionality $\hat{T}_{\rm lum}$ in the scaling 
relation
\begin{equation}
 T_{\rm lum} = \hat{T}_{\rm lum} \, \frac {(10 \, GH)^{2/3} \mu m_p} {2}
                             M_{200}^{2/3} \; \; .
\end{equation}
Values of $\hat{T}_{\rm lum}$ from simulations in the literature
can differ by as much as 30\%; examples include $\hat{T}_{\rm lum}
\approx 0.9$ from Evrard \etal (1996), $\hat{T}_{\rm lum}
\approx 0.8$ from Bryan \& Norman (1998), $\hat{T}_{\rm lum}
\approx 0.9$ from Thomas \etal (2001), and $\hat{T}_{\rm lum}
\approx 0.6$ from the non-radiative simulations of Muanwong \etal (2001).
The vast majority of these simulations ignore radiative cooling 
and supernova heating, both of which can elevate $\hat{T}_{\rm lum}$.
Mass-temperature relations derived from spatially-resolved
X-ray observations now suggest that the normalization constant
$\hat{T}_{\rm lum}$ derived from simulations is indeed too low
(Horner \etal 1999; Nevalainen \etal 2000; Finoguenov \etal 2001).
Furthermore, cluster-formation simulations that include
radiative cooling also yield a higher normalization constant;
for example, $\hat{T}_{\rm lum} \approx 1.1$ in the radiative 
simulations of Muanwong \etal (2001).  

Our modified-entropy models support these findings.
Figure~\ref{mtrel} compares two $M_{500}$-$T_{\rm lum}$ relations
derived from our modified-entropy models with the data from
Finogeunov \etal (2001) and Nevalainen \etal (2000).  In these
relations, we scale to $M_{500}$, the halo mass inside the radius

% ------------------ fig -----------------
\vspace{\baselineskip}
%\begin{figure}
\epsfxsize=3in 
\centerline{\epsfbox{f21.epsi}}
\figcaption{\footnotesize
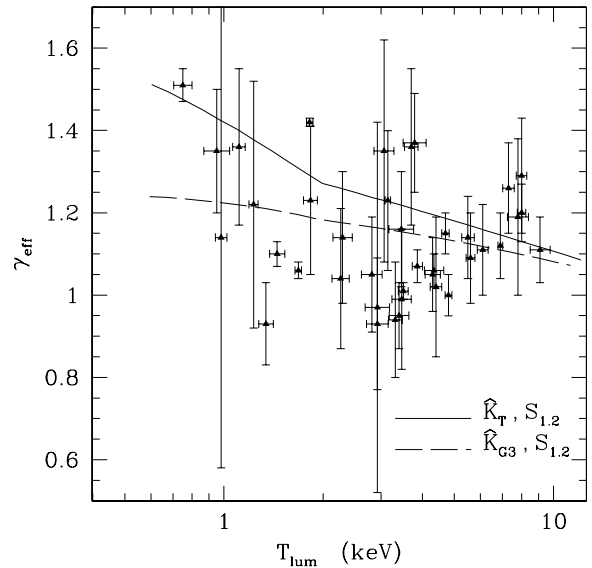
Effective polytropic index $\gamma_{\rm eff}$ as a function of 
luminosity-weighted temperature $T_{\rm lum}$.  The triangles
show $\gamma_{\rm eff}$ values derived from observed clusters
by Finoguenov \etal (2001).  The lines show values of
$\gamma_{eff}$ at $r/r_{200} = 0.2$ in truncated modified-entropy
models ($\hat{K}_T$, solid line) and radiative-loss models 
($\hat{K}_{G3}$, dashed line).
\label{gammas}}
\vspace{\baselineskip}
%\end{figure}
% ----------------------------------------

\vspace*{3em}

% ------------------ fig -----------------
\vspace{\baselineskip}
%\begin{figure}
\epsfxsize=3in 
\centerline{\epsfbox{f22.epsi}}
\figcaption{\footnotesize
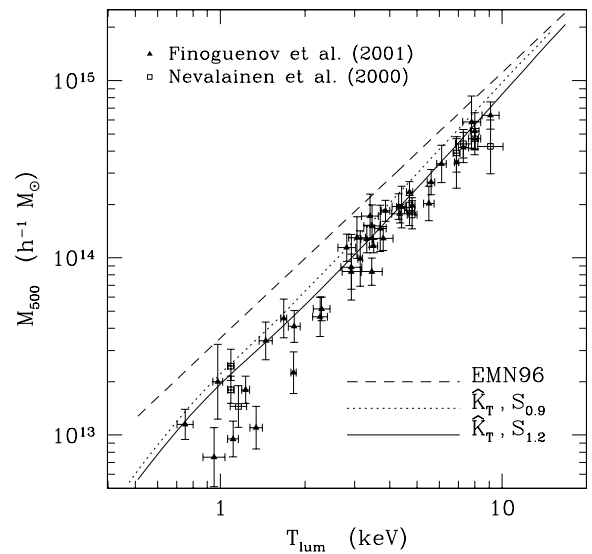
Relation between $M_{500}$ and luminosity-weighted temperature
($T_{\rm lum}$).  Solid triangles show mass measurements by
Finoguenov etal (2001), using $\beta$-model fitting and
the assumption of hydrostatic equilibrium.  Open squares
show mass measurements by Nevalainen \etal (2000), also
using $\beta$-model fitting and hydrostatic equilibrium.
Both of these data sets were corrected for the presence
of cooling flows, so we compare them to our models without
radiative losses.  The solid and dotted lines show 
the $M_{500}$-$T_{\rm lum}$ relations derived from $\hat{K}_T$ 
models using the concentration parameter relations $S_{1.2}$ 
and $S_{0.9}$, respectively (see Figure~\ref{c200}).  
The corresponding relations derived from shifted models
($\hat{K}_S$) are virtually identical.
The dashed line shows the $M_{500}$-$T_{\rm lum}$ relation
derived by Evrard \etal (1996) from simulations that do
not include radiative cooling or supernova feedback.
\label{mtrel}}
\vspace{\baselineskip}
%\end{figure}
% ----------------------------------------

\clearpage

\noindent
$r_{500} $ within which the mean matter density is $500 \rho_{cr}$,
because measurements of $M_{500}$ are thought to be more 
reliable than measurements of $M_{200}$.  The data have been
corrected for any cooling flow that might be present in the
cluster core, so we compare them to our truncated models
with concentration laws $S_{0.9}$ and $S_{1.2}$.  The
corresponding $M_{500}$-$T_{\rm lum}$ relations generated from
shifted models are virtually identical.  Both the truncated
and shifted models agree well with the data, but those based 
on concentration relation $S_{1.2}$ appear to match the data 
slightly better.

This agreement between the observed $M_{500}$-$T_{\rm lum}$
relation and the modified-entropy models is quite compelling,
considering that the models contain no adjustable parameters.
For comparision, Figure~\ref{mtrel} also shows the relation
derived from simulations by Evrard \etal (1996), which has
a lower temperature normalization (equivalent to a higher mass
normalization) and a shallower slope.  The steeper slope
of the $M_{500}$-$T$ relation from modified-entropy models 
arises primarily from the dependence of $\hat{T}_{\rm lum}$ on halo
concentration.  The higher concentration of low-mass
halos leads to a higher luminosity-weighted temperature
for a given mass.  Komatsu \& Seljak (2001) find a similar
effect in their polytropic cluster models.  Entropy modification
produces some additional steeping in the relation and shifts
its normalization to higher temperatures (see Figure~\ref{tlums_ts}).

The greater dispersion and offset toward lower masses at
$T_{\rm lum} \approx 1$ seen in Figure~\ref{mtrel} could be 
due to surface-brightness bias (Lloyd-Davies, Bower, \&
Ponman 2002).  Because the gas temperature
of intracluster gas in our models tends to decline with radius 
at large radii, the luminosity-weighted temperature depends 
on the maximum radius over which that temperature is
measured.  This effect is negligible in clusters because the
bulk of the emission is concentrated around the cluster's core,
but it can be significant in groups.  Figure~\ref{mtrel_t_sb}
shows how the observed $M_{500}$-$T_{\rm lum}$ relation would 
depend on the limiting surface-brightness for truncated models
with $S_{1.2}$ concentrations.  Increasing the surface brightness
threshold changes the derived relation in the sense that is observed.
Figure~\ref{mtrel_sb5e15} shows the relations for different entropy
modification schemes at a fixed surface-brightness threshold of
$5 \times 10^{-15} \, \sbm$.  Truncated and shifted models are affected
similarly, but surface-brightness bias has less effect on the
radiative-loss model ($\hat{K}_{\rm G3}$) because its central temperature
gradient is closer to isothermal.  Some of the scatter to higher
temperatures may also arise from the scatter in concentrations 
expected in low-mass halos (e.g., Afshordi \& Cen 2002), but the 
analysis of Lloyd-Davies \etal (2002) suggests that 
halo concentration alone cannot account for the large dispersion
of the mass-temperature relation for groups.

Despite the excellent agreement between our models and the 
observations, a word of caution is in order.
Both the modified-entropy models and the X-ray mass derivations
assume that clusters are in hydrostatic equilibrium.  Simulations,
on the other hand, suggest the intracluster velocity field
is not completely relaxed, with turbulent velocities 
$\sim 10-20$\% of the sound speed (Ricker \& Sarazin 2001).
Furthermore, the luminosity-weighted temperature derived from
theoretical modeling is not necessarily identical to the
spectral-fit temperature derived from observations (Mathiesen
\& Evrard 2001).  Nevertheless, our results strongly imply that
entropy modification and the cosmology-dependent relationship 
between halo mass and concentration must be accounted for in 
simulations designed to normalize the cluster mass-temperature 
relation.

% ------------------ fig -----------------
\vspace{\baselineskip}
%\begin{figure}
\epsfxsize=3in 
\centerline{\epsfbox{f23.epsi}}
\figcaption{\footnotesize
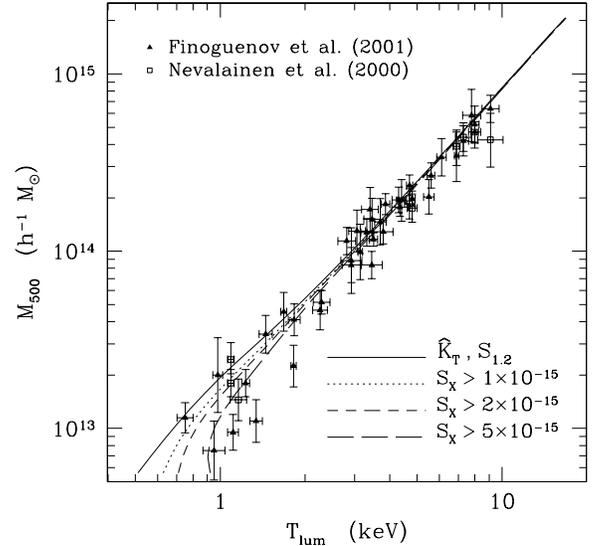
Relation between $M_{500}$ and luminosity-weighted temperature
($T_{\rm lum}$) at different surface-brightness levels.  Because
temperature declines with radius at large distances from the center
of our group models, the limiting surface brightness affects the 
observed mass-temperature relation.  The solid line shows the
$M_{500}$-$T_{\rm lum}$ relation for model $\hat{K}_T,S_{1.2}$.
The other lines show how that relation changes when
temperature is measured within regions where the bolometric
surface brightness in $\sbm$ exceeds $1 \times 10^{-15}$ (dotted), 
$2 \times 10^{-15}$ (short-dashed), and $5 \times 10^{-15}$ (long-dashed).
The data are the same as in Figure~\ref{mtrel}.  Some of the observed
dispersion below the solid line could stem from this surface-brightness
bias.
\label{mtrel_t_sb}}
\vspace{\baselineskip}
%\end{figure}
% ----------------------------------------

% ------------------ fig -----------------
\vspace{\baselineskip}
%\begin{figure}
\epsfxsize=3in 
\centerline{\epsfbox{f24.epsi}}
\figcaption{\footnotesize
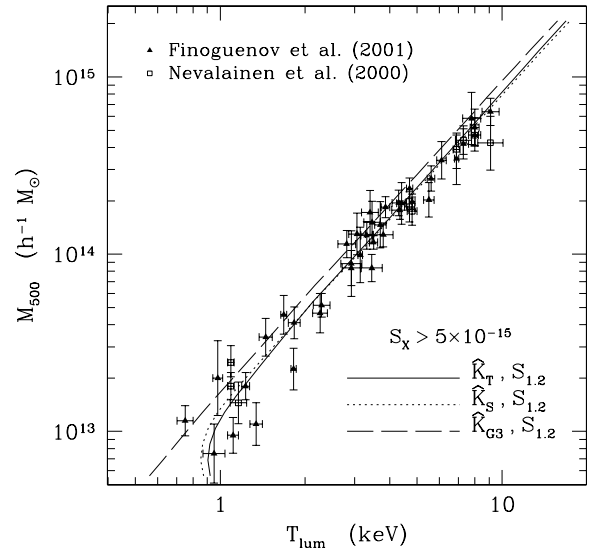
Relation between $M_{500}$ and luminosity-weighted temperature
($T_{\rm lum}$) above a surface brightness of $5 \times 10^{-15} \,
\sbm$.  The solid and dotted lines show the relation for truncated
and shifted models, respectively.  The dashed line shows the relation
for radiative-loss model $\hat{K}_{\rm G3}$.  The data are the same as 
in Figure~\ref{mtrel}.  The presence of gas below the cooling threshold 
near the center of the radiative-loss model lowers the central 
temperature, thereby mitigating the effects of the surface-brightness 
threshold.  
\label{mtrel_sb5e15}}
\vspace{\baselineskip}
%\end{figure}
% ----------------------------------------

\subsection{Luminosity-Temperature Relation}

Our previous work on modified-entropy models and the cooling
threshold demonstrated how well these models account for
the observed slope and normalization of the cluster 
luminosity-temperature relation (Voit \& Bryan 2001).
That paper compared truncated and shifted modified-entropy
models to cluster data that were corrected for the effects
of cooling flows.  Here, we revisit those models and
show how the $L_X$-$T_{\rm lum}$ relation changes when
gas below the cooling threshold is included.

Figure~\ref{ltrel} shows the $L_X$-$T_{\rm lum}$ relation
for clusters without cooling flows.  Of the three models 
shown, the truncated model with concentration relation
$S_{1.2}$ best fits the data, but all three models 
faithfully describe the $L_X$-$T_{\rm lum}$ relation
equally well on cluster scales.  Substituting concentration
relation $S_{0.9}$ shifts the models to slightly lower
temperatures, but the agreement with groups is almost
as good.  Using shifted instead of truncated entropy 
distributions elevates the X-ray luminosity on group
scales by a factor $\sim 2$.  

While there are essentially no free parameters in these models,
the cluster luminosity does depend on our assumed value of $\Omega_M$.
Raising $\Omega_M$ would lower the global ratio of baryons to 
dark matter, thereby lowering the luminosities of clusters in 
our models $\propto \Omega_M^{-2}$ (see Eq.~\ref{eq-lx}).  
Conversely, lowering $\Omega_M$ would raise the cluster
luminosities.  The excellent agreement found here for
$\Omega_M = 0.33$ should therefore reinforce confidence
in similar measurements of the matter density from the
baryon-to-dark-matter ratios of clusters (e.g., Evrard 1997).
Our models imply that groups are less suitable for such
measurements because their baryon-to-dark-matter ratio
within $r_{200}$ may be $\lesssim 50$\% of the global
value (see Figure~\ref{rmax}).

The tendency of the data at $\sim 1 \, \keV$ to scatter toward
the low-$L_X$, high-$T_{\rm lum}$ side of the model relations may 
be due to the same effects that cause similar 
scatter at those temperatures in the observed $M_{500}$-$T_X$
relation.  Figure~\ref{ltrel_t_sb} shows how surface-brightness 
bias affects the luminosity-temperature relations derived 
from our models.  Raising the surface-brightness threshold
within which luminosity and temperature are measured lowers 
$L_X$ while raising $T_{\rm lum}$, producing the same trend
seen in the data.  Measuring both of these quantities only
within regions where $S_X > 5 \times 10^{-15} \, \sbm$
produces relations that track the lower envelope of the
group data.  Unlike the cluster measurements, the data on
groups have not been corrected for the effects of gas below the
cooling threshold, which could counterbalance the surface-brightness
bias by raising $L_X$ and lowering $T_{\rm lum}$ (see Figure
\ref{ltrel_sb5e15}).  We therefore suspect that the dispersion
in the $L$-$T$ data of Helsdon \& Ponman (2000) arises from a
combination of differing surface-brightness biases and differing
amounts of gas below the cooling threshold.

Reproducing the behavior of clusters and groups with significant 
amounts of gas below the cooling threshold requires an additional  
parameter.  In our models, that parameter is $\alpha$, which
characterizes the slope of the entropy distribution function
below the cooling threshold. 
Figure~\ref{lt_mark} shows the relationship between $T_{\rm lum}$
and $L_{\rm ROSAT}$, the ROSAT-band (0.1-2.4 keV) luminosity,
in the data from Markevitch (1998) and in models with differing
values of ~$\alpha$.  ~Triangles ~show data corrected to remove ~the

% ------------------ fig -----------------
\vspace{\baselineskip}
%\begin{figure}
\epsfxsize=3in 
\centerline{\epsfbox{f25.epsi}}
\figcaption{\footnotesize
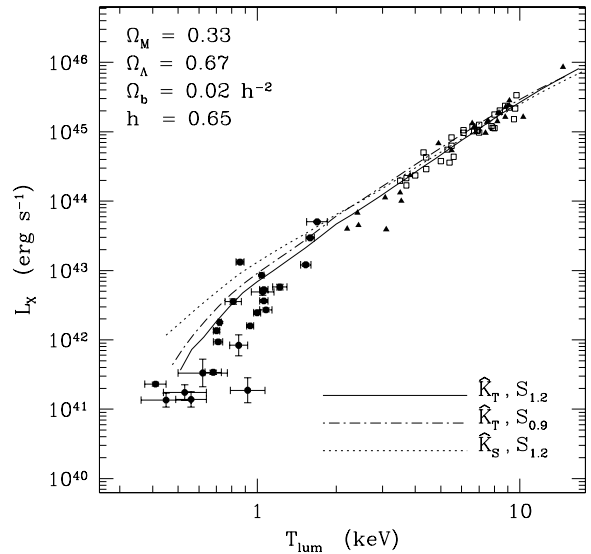
Relation between bolometric X-ray luminosity $L_X$ and 
luminosity-weighted temperature ($T_{\rm lum}$).  Solid triangles 
show measurements of clusters with insignificant cooling flows
compiled by Arnaud \& Evrard (1999).  Open squares show
cooling-flow corrected measurements by Markevitch (1998).
Solid circles show group data from Helsdon \& Ponman (2000).
Because the cluster data are not strongly influenced by cooling
flows, we compare them to models without radiative losses.
The solid and dotted lines show the $L_X$-$T_{\rm lum}$ relations 
derived from truncated ($\hat{K}_{\rm T}$) and shifted 
($\hat{K}_{\rm S}$) models, respectively, and the concentration
parameter relation $S_{1.2}$ (see Figure~\ref{c200}).  The
dot-dashed line shows the relation derived from model $\hat{K}_T$
using the concentration parameter relation $S_{0.9}$.
The models assume a standard $\Lambda$CDM cosmology with
$\Omega_M = 0.33$, $\Omega_\Lambda = 0.67$, and $\Omega_b
= 0.02 \, h^{-2}$, and a Hubble parameter of $h = 0.65$ has
been applied to both the models and the data.
\label{ltrel}}
\vspace{\baselineskip}
%\end{figure}
% ----------------------------------------

% ------------------ fig -----------------
\vspace{\baselineskip}
%\begin{figure}
\epsfxsize=3in 
\centerline{\epsfbox{f26.epsi}}
\figcaption{\footnotesize
Relation between $L_X$ and luminosity-weighted temperature
($T_{\rm lum}$) at different surface-brightness levels.  Because
temperature declines with radius at large distances from the center
of our group models, the limiting surface brightness affects the 
observed mass-temperature relation.  The solid line shows the
$L_X$-$T_{\rm lum}$ relation for model $\hat{K}_T,S_{1.2}$.
The other lines show how that relation changes when temperature 
and luminosity are measured within regions where the bolometric
surface brightness in $\sbm$ exceeds $1 \times 10^{-15}$ (dotted), 
$2 \times 10^{-15}$ (short-dashed), and $5 \times 10^{-15}$ (long-dashed).
The data are the same as in Figure~\ref{ltrel}.  Some of the observed
dispersion below the solid line could stem from this surface-brightness
bias.
\label{ltrel_t_sb}}
\vspace{\baselineskip}
%\end{figure}
% ----------------------------------------

% ------------------ fig -----------------
\vspace{\baselineskip}
%\begin{figure}
\epsfxsize=3in 
\centerline{\epsfbox{f27.epsi}}
\figcaption{\footnotesize
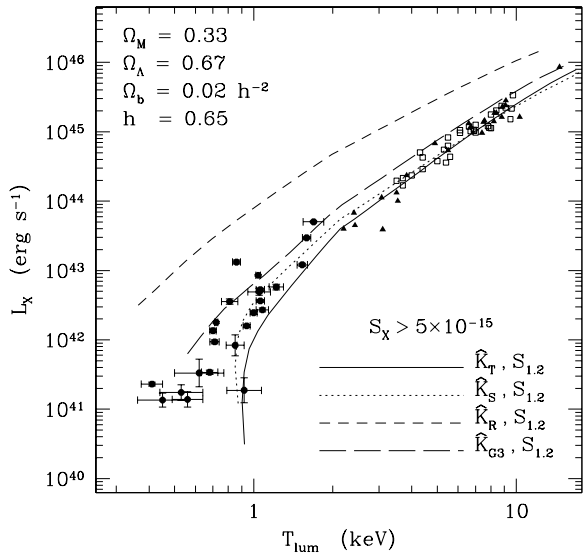
Relation between $L_X$ and luminosity-weighted temperature
($T_{\rm lum}$) above a surface-brightness threshold
of $5 \times 10^{-15} \,
\sbm$.  The solid and dotted lines show the relation for truncated
and shifted models, respectively.  The short-dashed and long-dashed
lines show the relations for radiative-loss models $\hat{K}_{\rm R}$
and $\hat{K}_{\rm G3}$, respectively.  The data are the same as 
in Figure~\ref{ltrel}.  The presence of gas below the cooling threshold 
near the centers of the radiative-loss models lowers the central 
temperature and raises the central luminosity.  Much of the dispersion
to the upper left of the solid and dotted lines near $\sim 1 \, \keV$
could be due to cooling flows, which have not been corrected for in
the group data.
\label{ltrel_sb5e15}}
\vspace{\baselineskip}
%\end{figure}
% ----------------------------------------

\vspace*{3em}

% ------------------ fig -----------------
\vspace{\baselineskip}
%\begin{figure}
\epsfxsize=3in 
\centerline{\epsfbox{f28.epsi}}
\figcaption{\footnotesize
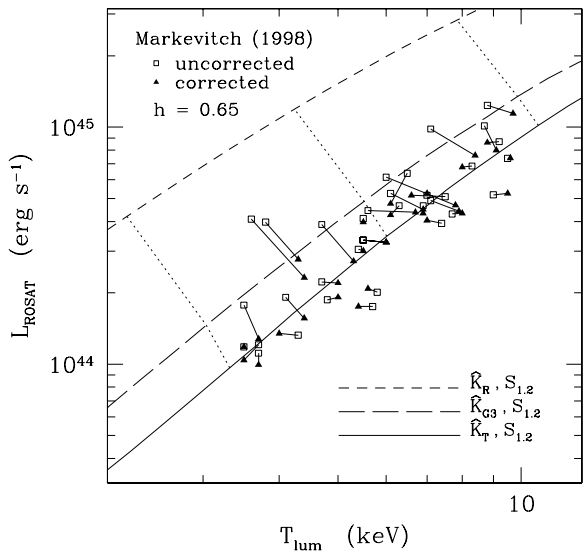
Relation between ROSAT ($0.1-2.4 \, \keV$) X-ray luminosity 
$L_{\rm ROSAT}$ and luminosity-weighted temperature ($T_{\rm lum}$).  
Open squares show uncorrected measurements from Markevitch
(1998), and solid triangles connected to those squares show
results for those same clusters after correcting for a central 
cooling flow.  (Error bars representing temperature uncertainty
$\sim 0.5 \, \keV$ have been suppressed for legibility.)
The solid line shows the $L_{\rm ROSAT}$-$T_{\rm lum}$
relation for truncated models, the long-dashed line depicts
model $\hat{K}_{G3}$ ($\alpha=3$), and the short-dashed line 
depicts model $\hat{K}_R$ ($\alpha=3/2$).  Dotted lines connect 
models for identical halo masses, showing how the 
$L_{\rm ROSAT}$-$T_{\rm lum}$ relation for a given halo mass shifts
to higher luminosity and lower temperature as $\alpha$ declines.
\label{lt_mark}}
\vspace{\baselineskip}
%\end{figure}
% ----------------------------------------

\noindent
effects of gas below the cooling threshold, and the squares 
connected to those triangles show the uncorrected data.  While 
the corrected data cluster around the line representing truncated 
models, many of the uncorrected data points lie closer to
the line representing radiative-loss models with $\alpha = 3$.  
A few uncorrected data points stray even farther from the truncated 
models, toward models with $\alpha < 3$.

Notice that the lines connecting data points with the largest
corrections are generally parallel to the lines connecting
modified-entropy models with identical halo mass.  This agreement
suggests that the parameter $\alpha$ adequately characterizes
the offset in $L_X$-$T_{\rm lum}$ space owing to intracluster
gas below the cooling threshold.  An interesting test of these
radiative-loss models would be to compare the value of $\alpha$
inferred from the luminosity-temperature offset to that implied
by the inner temperature gradient.

\subsection{What governs $\alpha$?}

The modified-entropy models we have constructed suggest that
most of the observable properties of clusters depend on only
two parameters.  The halo mass $M_{200}$ determines a sort
of cluster ``main sequence'' with well-defined $M_{500}$-$T_{\rm lum}$
and $L_X$-$T_{\rm lum}$ relations.  Deviations from those relations
and the severity of the inner temperature gradient both depend 
on a second parameter, $\alpha$, related to the amount of 
intracluster gas below the cooling threshold, but what physical
processes determine the value of $\alpha$?  

One is tempted to interpret $\alpha$ in terms of a mass cooling 
rate $\dot{M}_X$; however, {\em XMM-Newton} observations show little 
evidence for the low-temperature line emission expected in the 
standard cooling-flow picture (Peterson \etal 2001).  Thus, 
instead of viewing $\alpha$ as simply a measure of cooling-flow 
strength, we would like to suggest a wider range of possibilities:
\begin{itemize}
\item Entropy history.  Our simplistic schemes for entropy
      modification treat clusters as if they were assembled
      very early in time, with some standard initial entropy
      distribution.  In fact, the entropy history of the
      intracluster medium is likely to be far more complex,
      influenced in differing ways at various times by
      merger shocks, radiative cooling, and feedback.
      The dispersion seen in $\alpha$ could stem from
      differences in the entropy history of clusters,
      with some evolutionary paths leading to abundant
      gas below the cooling threshold and other paths
      leading to very little.  If this is the case, then
      a cluster's value of $\alpha$ would be an important
      window into its past.
\item Incomplete relaxation.  {\em Chandra} observations of
      several clusters have revealed ``cold fronts'' thought
      to be evidence of low-entropy gas sloshing around in
      the cores of those clusters (Markevitch \etal 2000; 
      Vikhlinin, Markevitch, \& Murray 2001).  Our models
      assume that the lowest-entropy gas in a cluster has
      already settled into hydrostatic equilibrium at the
      cluster's center.  Some of the apparent dispersion 
      in $\alpha$ could be due to differences in the degree 
      to which low-entropy gas has settled within cluster 
      cores following the last major merger.  \clearpage
\item Recent feedback.  Cooling and condensation of low-entropy
      gas at the center of a cluster is likely to lead to
      star formation and perhaps enhanced activity
      in the nucleus of the central galaxy (Fabian 1994).  Both
      star formation and radio jets can provide feedback,
      imparting entropy to the cluster core that inhibits 
      further condensation (B\"ohringer \etal 2002).  
      Perhaps these feedback processes control the value of $\alpha$.
\item Thermal conduction.  The presence of significant temperature 
      gradients within the cores of some clusters has often been
      cited as evidence that magnetic fields strongly suppress
      electron thermal conduction in clusters (Fabian 1994).  Recent
      {\em Chandra} observations of some particularly sharp 
      cold fronts have reinforced that supposition (Ettori \& Fabian
      2000).  However, these observations constrain 
      conduction in only one dimension and do not necessarily 
      rule out conduction in the other two dimensions.
      Furthermore, recent theoretical work suggests that
      magnetic fields might not suppress conduction as effectively
      as previously believed (Malyshkin \& Kulsrud 2001;
      Malyshkin 2001; Narayan \& Medvedev 2001).
      It is possible that the magnetic field geometry in
      a cluster's core places a lower limit on the value 
      of $\alpha$.
\end{itemize}

Investigating all of these possibilities is beyond the scope 
of this paper, but in the next section we explore how one
might self-consistently calculate the intracluster entropy
distribution within a hierarchical merging scenario.

\section{Coupling Entropy, Evolution, and Feedback}

We have shown that some very simple prescriptions for the
entropy distribution in present-day clusters lead to a
remarkably realistic set of cluster models.  However, the
astrophysics that leads to these entropy distributions is 
certainly more complicated than we have assumed.  Understanding
how the the physics of structure formation, cooling, and
feedback determine the intracluster entropy distribution
will require a more complete theory based on hierarchical
merging.  This section outlines a few of the basic issues 
involved in constructing such a theory.
 
\subsection{Entropy Histories}

Considering the entropy history of a single gas parcel 
affords some insight into how radiative cooling 
establishes an entropy threshold 
(Voit \& Bryan 2001).  Suppose that the parcel begins 
with entropy $K_1$ in a halo of temperature $T_1$ at some 
early time $t_1$.   In a hierarchical merging scenario, 
the parcel's halo will eventually merge with another
halo at time $t_2$.  Merger shocks will then raise
the parcel's entropy to some new value $K_2$, and
as that parcel settles within the new halo, it will
approach a new temperature $T_2$, similar to the
characteristic temperature of that new halo.  
Each subsequent merger will also raise
the parcel's temperature and entropy.  Thus, the
trajectory of a parcel through the entropy-temperature
plane can be schematically represented by a sequence
of points like that in Figure~\ref{entprop}.

The dashed lines in Figure~\ref{entprop} show how the
entropy threshold associated with cooling rises with time 
until it achieves its present value, indicated by the
solid line.  As long as a parcel's trajectory through the 
entropy-temperature plane remains above the cooling threshold, 
then radiative cooling and subsequent feedback will not 
significantly affect that parcel's entropy.  However,
any parcel that spends a large fraction of time below
that threshold will be subject to substantial entropy
loss, condensation, and whatever feedback ensues.
Feedback can continue as long as there are gas parcels
below the cooling threshold, but once cooling and feedback
have eliminated all gas below the cooling threshold,
feedback must cease.  Thus, the threshold for entropy 
modification established by both cooling and feedback 
corresponds to the cooling threshold defined by the
Hubble time $t_H$.
 
Although cooling sets the threshold for entropy modification,
the model we are describing is not a cooling-flow model.
Hierarchical structure formation will ensure that mergers 
disperse the products of condensation throughout the 
final cluster.  Stars and galaxies form out of the 
lowest-entropy gas long before the cluster itself forms.  
Hence, much of the low-entropy gas that would eventually
find its way into the cluster core in a simulation with 
no radiative cooling decouples from the intracluster medium 
before it can participate in a centrally-focused cooling flow.  
Finally, for reasons we will describe next, a certain amount
of feedback is needed to prevent the incipient intracluster
medium from overcooling at high redshift. 

\vspace*{3em}

% ------------------ fig -----------------
\vspace{\baselineskip}
%\begin{figure}
\epsfxsize=3in 
\centerline{\epsfbox{f29.epsi}}
\figcaption{\footnotesize
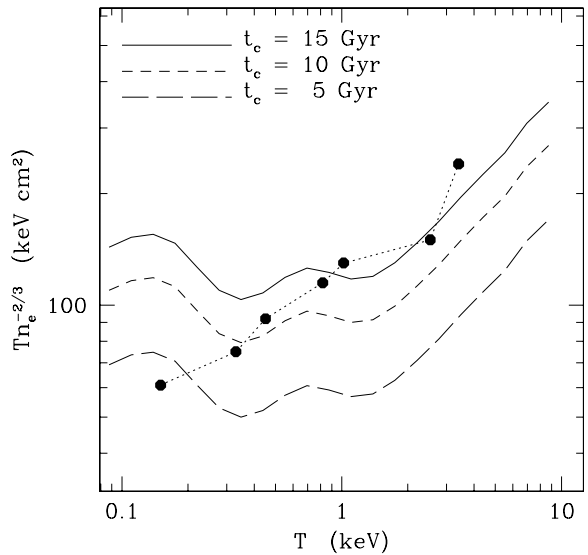
Entropy-temperature diagram showing the entropy history
of a gas parcel and the evolution of the cooling threshold
with time.  Solid and dashed lines show the locus in the
$T$-$Tn_e^{-2/3}$ plane at which the cooling time $t_c$ of
a gas parcel equals 15~Gyr (solid line), 10~Gyr (short-dashed
line) and 5~Gyr (long-dashed line).  This locus moves upward
through the diagram as the universe ages.  The dots connected
by dotted lines schematically illustrate the entropy history
of a gas parcel as described in the text.  As long as the
parcel's trajectory through the entropy-temperature plane remains
above the cooling threshold, it will not substantially cool.
However, both cooling and feedback triggered by that cooling
will inevitably modify the entropy of gas parcels that find 
themselves below the cooling threshold.
\label{entprop}}
\vspace{\baselineskip}
%\end{figure}
% ----------------------------------------

\clearpage
\subsection{Overcoming Overcooling}
\label{overcool}

Overcooling is one of the classic problems that models
of hierarchical structure formation must overcome
(Cole 1991; White \& Frenk 1991; Blanchard, Valls-Gabaud, \&
Mamon 1992; Balogh \etal 2001).  If radiative cooling were
allowed to proceed unchecked as structure formed, then 
a substantial percentage ($\gtrsim 20$\%) of the 
universe's baryons would have been locked into condensed objects
before the era of cluster formation.  Because the fraction
of uncondensed baryons in present-day clusters is $\sim 90$\%, feedback
processes like supernova heating are presumed to inhibit cooling and
condensation of high-redshift baryons (White \& Rees 1978; Cole 1991;
White \& Frenk 1991).

In our models, overcooling manifests itself as an entropy 
threshold $K_c$ that is much higher than the characteristic
entropy of an unmodified high-redshift halo.  To illustrate 
this effect, let us assume that the progenitor halo of a 
present-day cluster has a temperature $T(z)$ that depends 
on redshift.  The entropy threshold for that halo is then 
$K_c(z) \propto \{ T^{1/2}(z) \, \Lambda[T(z)] \, t_H(z) \}^{2/3}$, and 
the characteristic entropy within $r_{200}$ in the unmodified halo is 
$K_{200}(z) \propto T(z) H^{-4/3}$.  The ratio of these 
two quantities in the high-redshift limit is thus
\begin{equation}
  \frac {K_c} {K_{200}}  \propto  \left\{ \frac  
                                  {\Lambda [T(z)]} {T(z)} \right\}^{2/3}
                                  [H(z) t_H(z)]^{2/3} H^{2/3}(z) \; \; .
\end{equation}
Because $H(z)$ rises and $T(z)$ declines with increasing 
redshift, this ratio must exceed unity at early times.

We can use extended Press-Schechter theory (Bond \etal 1991; 
Bower 1991) to approximate $T(z)$ and estimate the
temperature, entropy, and redshift scales at which overcooling
is problematic.  First, to demonstrate the principle with simple 
scaling relations, we consider a SCDM ($\Omega_M = 1$) universe with 
a power spectrum given by an $n=-2$ power law.  
In that case, the characteristic mass scale evolves like 
$M \propto (1+z)^{-6}$, and assuming that $M \propto T^{3/2}
(1+z)^{-3/2}$ yields $T(z) \propto (1+z)^{-3}$.  
At low redshifts, where $T$ is high, we have $\Lambda \propto 
T^{1/2}$ and $K_c / K_{200} \propto  (1+z)^2$.
However, at high redshifts, when the progenitor temperature 
is smaller, we have $\Lambda \propto T^{-1/2}$ and $K_c / K_{200}
\propto (1+z)^4$.  Thus, the progenitor of a halo with 
$K_c / K_{200} \approx 0.1$ at $z \approx 0$ would have
$K_c / K_{200} \sim 1$ at $z \sim 1-2$.

Figure~\ref{ent_evol} illustrates this effect more precisely
for a $\Lambda$CDM universe seeded by a power spectrum 
with shape parameter $\Gamma=0.17$.  The dashed line shows 
$K_{200}(z)$ for a halo that reaches $T = 5 \, \keV$ at $z=0$.  
The solid line shows 
the entropy threshold $K_c$ in that evolving 
halo for a metallicity of one-third solar.  Feedback must impart 
an entropy $\sim 30 \, \keV \, {\rm cm}^2$ to prevent overcooling
at $z \gtrsim 3$.

\subsection{Interplay between Feedback and Merging}
\label{interplay}

Feedback can solve the overcooling problem, but it introduces 
another, more subtle problem into our framework.  Our 
modified-entropy models assume that gas above the cooling threshold
follows the unmodified distribution $K_0(M_g)$.  However,
the need for feedback at early times implies that a large
proportion of intracluster gas has experienced
entropy modification at least once in its history.  Somehow,
the process of cluster formation needs to restore the unmodified
distribution to gas above the threshold at $z=0$.

% ------------------ fig -----------------
\vspace{\baselineskip}
%\begin{figure}
\epsfxsize=3in 
\centerline{\epsfbox{f30.epsi}}
\figcaption{\footnotesize
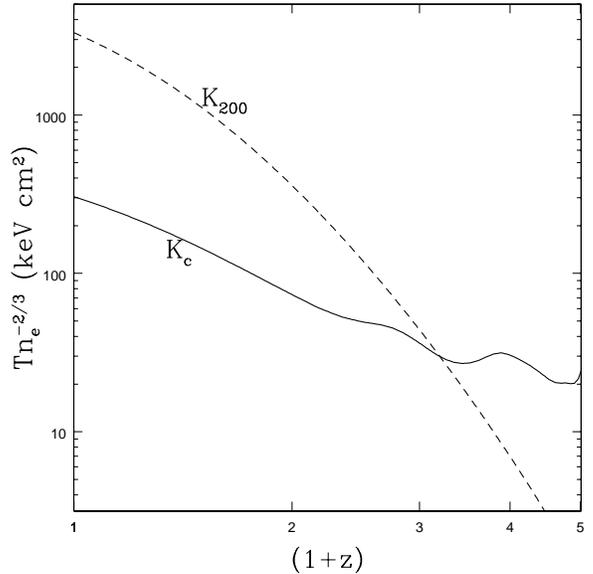
Entropy evolution with redshift.  The dashed line shows
how the characteristic entropy $K_{200}$ of the progenitor halo
of a $5 \, \keV$ cluster at $z=0$ evolves with redshift in
a $\Lambda$CDM universe with shape parameter $\Gamma = 0.17$.
The solid line shows how the cooling threshold $K_c$ associated
with that halo depends on redshift.  Because $K_{200}$ falls
below the cooling threshold at $z \gtrsim 3$, feedback is
needed to prevent the majority of the baryons associated
with the progenitor from condensing and forming stars.
\label{ent_evol}}
\vspace{\baselineskip}
%\end{figure}
% ----------------------------------------

\vspace*{1em}

Exactly how that happens is unclear.  Consider what happens
in a merger between two halos containing isothermal gas at
temperatures $T_1$ and $T_2$ that collide at velocity $v$.
Merger shocks with Mach number ${\cal M}_i \propto vT_i^{-1/2}$
will propagate through each halo's gas, raising the entropy
of that gas by some more-or-less uniform factor depending
on ${\cal M}_i$.  This mechanism could potentially amplify the
effects of early feedback, which must raise the intrahalo
entropy to at least $K_c$ to avoid overcooling.  Subsequent
shocks would then boost the entropy even further, along
a path parallel to the dashed line in Figure~\ref{ent_evol}.
If this mechanism dominates, then very little gas should
remain near $K_c$ at $z \sim 0$.  

However, early feedback is likely to diminish the effects 
of mergers, particularly in gas with low entropy.  If feedback
raises all the gas in the progenitor halo of a cluster to
$\sim K_c$, then the intrahalo medium will be nearly isentropic,
with high-temperature gas at the center of the halo and
lower-temperature gas at the outskirts (see Figure~\ref{tprofs}).
When mergers occur, Mach numbers associated with lower-entropy
gas in the halo's core will therefore be lower than those
associated with higher-entropy gas near the virial radius.
Hence, merger shocks will act to steepen any entropy gradient
that exists.  Dynamical friction may also add to this effect,
because the cores of merging halos tend to lose orbital
energy before they collide and merge.  Another effect to
consider is the quasi-continuous accretion of gas associated
with small halos, for which the Mach number is likely to
be quite high.  This type of accretion will enhance the
amount of high-entropy gas in the final cluster.  Thus, it
seems possible that the modified-entropy models we have
developed can successfully be linked with hierarchical
structure formation, although many details remain to
be worked out.

\subsection{Excess Energy}

Several recent analyses of similarity breaking in groups and
clusters have cast the problem in terms of energy rather than
entropy (e.g., Wu \etal 1998, 2000; Lloyd-Davies, Ponman, \& Cannon 
2000; Loewenstein 2000; Bower \etal 2001; Lloyd-Davies \etal 2002).  
For example, if one knows what the configuration of the intracluster 
gas would be in the absence of non-gravitational heating and cooling 
processes, then one can opt to define the ``excess energy'' of a 
cluster's gas to be the difference in total energy between
that unmodified configuration and the actual configuration.
Depending on what is assumed about the unmodified configuration,
these excess energies can range from $\sim 0.3-1.0 \, \keV$ for
groups (e.g., Lloyd-Davies \etal 2000, 2002) to over 2~keV for 
clusters (e.g., Wu \etal 2000).  

The results of this paper indicate that some combination of heating 
and cooling is responsible for similarity breaking because cooling 
is needed to explain the core entropies of clusters and groups and 
heating is needed to prevent overcooling at early times.  Therefore,
we prefer to compute excess energy in a way that explicitly accounts
for the separate contributions of cooling and heating.
First, we assume that the baseline state of a cluster of mass
$M_{200}$ is a truncated model from which cooling and condensation
has removed a fraction $f_*$ of the lowest-entropy gas.  
In other words, the initial configuration of the intracluster
medium has an entropy floor equal to $K_0(f_*)$ and contains 
a fraction $1-f_*$ of the cluster's baryons.  Then, we assume
that the final state of the cluster is a shifted model with
the same proportion of baryons but an entropy floor equal to 
the cooling threshold, so that it obeys the observed $M$-$T$ and
$L$-$T$ relations.  Formally, the entropy distribution of the
final state is $K(f_g) = K_0(f_g) + K_c(T_{\rm lum})$ with an 
outer boundary where $f_g = 1-f_*$. Because the amount of hot baryonic 
gas in both models is the same, the amount of heat input required to 
explain the cluster scaling relations after a fraction $f_*$ 
of the intracluster gas has condensed equals the energy input 
needed to convert the initial configuration to the final 
configuration, including work done at the outer boundary of 
the intracluster medium.

Figure~\ref{xs_energy} shows the energy difference between these
configurations as a function of halo mass for different values 
of the condensed baryon fraction $f_*$.  To facilitate comparisons 
with other treatments of excess energy, we divide this energy 
difference by the total number of particles in an unmodified
cluster ($f_b M_{200} / \mu m_p$), giving the quantity 
$\Delta \epsilon$ in units of keV per particle.  When the
condensed baryon fraction is small ($f_* = 0.02$), the
excess energy needed to account for the scaling relations of
massive clusters ($\sim 10^{15} \, M_\odot$) approaches the
$\sim 3 \, \keV$ level.  However, this large amount of excess
energy depends heavily on the assumed form of the unmodified entropy
distribution, which is likely to be a poor approximation for
these purposes.  The large difference in the excess energy 
of massive clusters between models with $f_* = 0.02$ and 
$f_* > 0.05$ indicates that a large amount of energy is needed
to lift the lowest-entropy gas out of the very center of the
gravitational potential.  Yet, Figure~\ref{denprofs} shows that
the unmodified intracluster density distribution is unlikely
to be as centrally condensed as we have assumed.  Even in the 
absence

% ------------------ fig -----------------
\vspace{\baselineskip}
%\begin{figure}
\epsfxsize=3in 
\centerline{\epsfbox{f31.epsi}}
\figcaption{\footnotesize
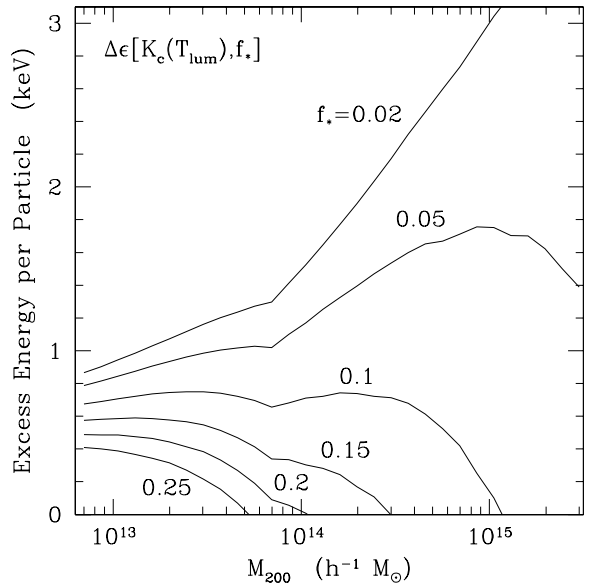
Excess energy required to overcome the cooling threshold for different
values of the condensed baryon fraction ($f_*$).  At each halo mass
$M_{200}$, the quantity $\Delta \epsilon [K_c(T_{\rm lum}),f_*]$
is the energy input per particle needed to transform the truncated 
entropy distribution $\hat{K}_{\rm T}(f_*)$, from which a fraction
$f_*$ of the lowest-entropy baryons have been removed, into a shifted 
distribution $\hat{K}_S = \hat{K}_0 + \hat{K}_c$ containing the
remaining fraction $1-f_*$ of the baryons.  This transformation
corresponds to first removing gas without adding heat energy, then
adding enough heat to reproduce the $L$-$T$ and $M$-$T$ relations.
The energy input required to produce these relations for plausible
condensed-baryon fractions of $f_* \approx 0.1-0.2$ is $\sim 0.4-0.7
\, {\rm \keV \, particle^{-1}}$ at group scales and is negligible
for hot clusters.
\label{xs_energy}}
\vspace{\baselineskip}
%\end{figure}
% ----------------------------------------

\noindent
 of cooling, the true unmodified profile probably 
corresponds more closely to a truncated model with $\hat{K}_c 
\sim 0.05-0.1$, which substantially reduces the excess energy 
burden.  Thus, when calculating excess energies for clusters, one
must make sure that the baseline model accurately represents
the cluster configuration in the absence of heating.

The curve with $f_* = 0.1$ is likely to be a more realistic
representation of the heat input needed to explain the scaling
properties of massive clusters for two reasons: (1) this value
of $f_*$ roughly corresponds to the fraction of cluster 
baryons that have condensed into stars, and (2) when 
$f_* \gtrsim 0.1$, truncation eliminates the innermost part 
of the entropy distribution that is discrepant with simulations.
At $10^{15} \, M_\odot$, this excess energy curve suggests
that only a fraction of a keV per particle is needed to explain cluster
scaling relations.  That is because a shifted model with
an entropy shift determined by the cooling threshold is
virtually identical to a truncated model with the innermost
$f_*$ of gas removed.  Likewise, the mass scales at which 
curves with larger values of $f_*$ intercept the zero-energy 
level also indicate cases in which the truncated and shifted
models are virtually identical, requiring no excess heating 
to produce the observed scaling relations.  In the limiting
case of zero supernova heating (e.g., Bryan 2000, Muanwong \etal
2001), these intercept points show the fraction of condensed baryons 
implied by the $L$-$T$ relation as a function of halo mass.

If the condensed baryon fractions of groups are similar to
those of clusters, then $\sim 0.3-0.7 \, \keV$ of excess
energy is needed to explain the $M$-$T$ and $L$-$T$ relations
at $\sim 1 \, \keV$.  The main difference between the initial
configuration and the final configuration of groups with
$f_* \sim 0.1-0.25$ is that the gas in the final configuration
is considerably more extended, with substantially more gravitational
potential energy.  Because models with higher $f_*$ contain less intragroup
gas, they are less extended, and less energy is needed to change
their configuration.  We note that the excess energy needs of 
groups are comparable to estimates of the supernova energy released
over the course of the group's history (e.g., Ponman \etal 1999,
Loewenstein 2000; Pipino \etal 2002), which may mean that no 
other source of energy is needed to explain their scaling
properties, if that energy efficiently heats the intragroup
medium.

The bottom line here is that calculations of the heat input
needed to account for the scaling relations of clusters and
groups depend critically on what is assumed to happen in 
the absence of heat input.  Most calculations of excess energy
have assumed that the intracluster gas density obeys either 
a simple polytropic equation of state (e.g., Wu \etal 2000; 
Loewenstein 2000) or a $\beta$-model density profile (e.g.,
Bower \etal 2001; Lloyd-Davies \etal 2002) whose parameters
are adjusted according to the amount of heat input.  The 
modified-entropy models discussed in this paper allow for
cooling and condensation of the lowest-entropy gas to occur
before heat is added.  Figure~\ref{xs_energy} demonstrates that the
amount of heating needed to explain the scaling relations
depends strongly on the amount of cooling that occurs prior 
to heat input and that realistic amounts of ``precooling'' 
substantially reduce the required amount of non-gravitational
heating.

\section{Conclusions}

A realistic family of models for clusters in hydrostatic
and convective equilibrium can be constructed using
the Navarro, Frenk, \& White (1997) density profile
and some rudimentary prescriptions for how radiative
cooling and feedback induced by that cooling modify 
the entropy distribution of intracluster gas.  The models 
presume that the intracluster entropy distribution in the 
absence of cooling and feedback would be identical to that 
of gas that takes on the same density distribution as the 
dark matter as it settles into the cluster's potential
well---an assumption supported by numerical simulations.
Because the lowest-entropy gas within that distribution
can cool within a Hubble time, its entropy must somehow
be modified.  The threshold entropy in our prescriptions
for entropy modification corresponds to the entropy at
which the cooling time of intracluster gas equals the
age of the universe.  The prescriptions themselves
include truncation of the entropy distribution at
the threshold entropy, shifting of the entropy distribution
by adding the threshold entropy to the entire distribution,
and a qualitative implementation of radiative cooling.

Exploring the properties of dimensionless models depending
only on the halo concentration $c$, the dimensionless entropy
threshold $\hat{K}_c$, and the prescription for entropy
modification reveals that: 
\begin{itemize}
\item Removal of low-entropy gas acts to flatten the core density 
      profile, regardless of whether cooling or heating eliminates
      that low-entropy gas (see Figures~\ref{denprofs}
      and \ref{denprofs_all}).
\item The observable properties of clusters depend more
      critically on the threshold entropy than on the 
      mode of entropy modification.  For example, truncation
      of the intracluster entropy distribution at a given value
      of $\hat{K}_c$ and shifting of the entropy distribution by 
      that same amount lead to very similar pressure, density, 
      and temperature profiles within the cluster's
      virial radius (see Figures~\ref{denprofs_all} and \ref{tprofs}). 
\item Within about 30\% of the virial radius, the surface-brightness
      profiles of truncated and shifted models are very similar
      to $\beta$-models (see Figure~\ref{sbright}).  Raising the 
      entropy threshold tends to lower the best-fitting value of 
      $\beta$, while raising the halo concentration tends to 
      decrease the core radius (see Figures~\ref{betafits_t}
      and \ref{betafits_s}).
\item All of our prescriptions for entropy modification tend
      to augment the luminosity-weighted temperature of a cluster
      as the entropy threshold rises.  However, that temperature
      increase is relatively small because the increased central
      entropy decreases the luminosity of high-temperature gas
      in the core.  Thus, modifiying the entropy of a cluster
      shifts the bulk of a cluster's luminosity to larger radii,
      where gas temperatures tend to be smaller, mitigating
      the effects of the entropy increase (see \S~\ref{emodtemp}).
\item All of our prescriptions for entropy modification reduce the 
      X-ray luminosity of a cluster as the entropy threshold rises.  In 
      the high-threshold limit, dimensionless luminosity scales as 
      $\hat{L} \propto \hat{K}_c^{-3/2}$, owing to the asymptotic $\rho 
      \propto r^{-3}$ scaling of the unmodified NFW density distribution 
      at large radius (see \S~\ref{emodlum}).
      When luminosity scales with the entropy threshold in this
      way, the $L$-$T$ relation should not flatten below $\sim 2 \,
      \keV$, where line cooling dominates free-free emission, and the 
      relation should also evolve little with time.  The observed $L$-$T$
      relation shares both of these features.  Furthermore, this
      scaling of luminosity with the entropy threshold leads to
      an $L$-$T$ relation slightly steeper than $L \propto T_{\rm lum}^{5/2}$,
      with the extra steepening coming from the tendency for low-temperature
      halos to be more concentrated, also in agreement with
      observations.
\end{itemize}

Adopting relations between halo concentration $c$ and halo mass
$M_{200}$ that are drawn from simulations and supported by observations
enables us to generate models for real clusters that reproduce
many of their observable properties:
\begin{itemize}
\item Fitting $\beta$-models to our modified-entropy clusters
      yields $\beta$ and core-radius values that
      are similar to those of observed clusters.  Our models
      reproduce the observed tendency for $\beta$ to be lower 
      in low-temperature halos, a behavior that arises because
      the entropy threshold determined by cooling has a much
      larger impact on low-temperature halos.  Our models
      also reproduce the observed relationship between core
      radius and temperature, implying that $r_c \approx 0.1 \,
      r_{200}$, regardless of temperature.  Despite the
      flattening of the density profile in low-temperature halos, 
      the core radius remains near $0.1 \, r_{200}$ because of
      the increased halo concentration (see Figures~\ref{betafits_t}
      and \ref{betafits_s}).  The large dispersion observed in
      $\beta$ and $r_c$ could arise from a number of effects,
      some involving observational systematics such as 
      surface-brightness bias and the range of radii in the
      fit, others involving physical differences such as
      cluster-to-cluster variations in the amount of gas
      below the cooling threshold (see Figures~\ref{bmods} and 
      \ref{bmods_all}).
\item The temperature gradient at small radii in our model
      clusters depends sensitively on the amount of gas
      below the cooling threshold, closely related to the
      parameter $\alpha$ in our radiative-loss models.  In
      our most extreme radiative-loss model, with $\alpha = 3/2$,
      the temperature gradient at $\lesssim 0.3 \, r_{200}$ is very
      similar to the ``universal'' temperature gradient observed
      by Allen \etal (2001).  As the parameter $\alpha$ rises
      to infinity, the limit in which no gas lies below
      the cooling threshold, the inner temperature gradient flattens 
      and then reverses, monotonically decreasing with radius
      in the large $\alpha$ limit (see Figure~\ref{tinner}).
\item All of our models have a negative temperature gradient 
      at large radii because the underlying potential is steeper
      than isothermal at those radii.  In order to compare our
      models with observations, we compute the effective polytropic
      index $\gamma_{\rm eff} \equiv d \ln P / d \ln \rho$ as a
      function of radius.  At the virial radius, we find $\gamma_{\rm eff}
      \approx 1.1 - 1.2$, but the behavior of $\gamma_{\rm eff}$ at
      smaller radii depends strongly on the nature of entropy
      modification.  In models with no gas below the cooling threshold,
      the cluster core is nearly isentropic, with $\gamma_{\rm eff} 
      \approx 5/3$.  In models with significant amounts of gas
      below that threshold, we find $\gamma_{\rm eff} \lesssim 1$ within the
      core radius, a consequence of the positive temperature gradient
      (see Figure~\ref{gamma_profs}).  Values of $\gamma_{\rm eff}$
      evaluated at $0.2 \, r_{200}$ agree with observations, within
      the large observational uncertainties (see Figure~\ref{gammas}).
\item The mass-temperature relations derived from our truncated and shifted 
      modified-entropy models, which have no gas below the cooling
      threshold, agree well with those derived from cluster observations
      that have been corrected for the presence of cooling flows
      (see Figure~\ref{mtrel}).  The slope of
      this relation is steeper than the $M \propto T^{3/2}$ expectation
      from self-similar scaling because low-temperature halos are more
      concentrated, leading to a slightly higher temperature for a given
      mass (see Figure~\ref{tlums_ts}), a feature these models share with
      those of Komatsu \& Seljak (2001).  Clusters of a given temperature
      are less massive than those in simulations without cooling and
      feedback because entropy modification owing to these processes
      shifts all of our model clusters to higher temperature.  However,
      the agreement between observations and our models does not 
      necessarily imply that this $M$-$T$ relation is without problems,
      as both the observational determinations and the models assume
      that the intracluster medium is an ideal gas in hydrostatic 
      equilibrium, which may not be the case.  Furthermore, this relation
      does not apply to clusters that have not been corrected for
      cooling flows---the temperature normalization at a given mass
      in our radiative-loss models can be up to 30\% lower, depending
      on the value of $\alpha$ (see Figures~\ref{tlums_r} and \ref{tlums_ta}).
\item The luminosity-temperature relations derived from truncated and
      shifted models also agree well with those derived from cooling-flow
      corrected observations of clusters (see Figure~\ref{ltrel}).
      While there are no adjustable parameters in these models, the
      baryon-to-dark-matter ratio does depend on our assumed value
      of $\Omega_M = 0.33$.  Thus, this agreement can be taken as
      additional evidence for a low value of $\Omega_M$.
\item Gas below the cooling threshold both raises the X-ray luminosity
      and lowers the luminosity-weighted temperature of our model clusters. 
      Hence, the $L$-$T$ relations derived from our radiative-loss models
      lie above and to the left of the truncated and shifted models
      in the $L$-$T$ plane.  Changing $\alpha$ leads to a displacement
      almost perpendicular to the $L$-$T$ relations themselves (see 
      Figure~\ref{lt_mark}).  Correcting for cooling-flow 
      emission leads to a very similar displacement in $L$-$T$ space, 
      suggesting that the $L$-$T$ relation for uncorrected clusters may 
      depend almost entirely on the two parameters $M_{200}$ 
      and $\alpha$.  It will be interesting to see whether values
      of $\alpha$ derived from displacement in $L$-$T$ space agree
      with those implied by the inner temperature gradient.
\end{itemize}
   
The modified-entropy models we have derived can account for many
of the observed properties of present-day clusters, but they are
obviously too simplistic to tell us how those clusters got to be
the way they are.  Somehow, hierarchical structure formation generated
intracluster entropy distributions similar to our simple prescriptions
for entropy modification.  Cooling, feedback, and merging all have
important roles to play.  The consistency of our models with the
data strongly suggests that radiative cooling sets the level of 
the intracluster entropy threshold and substantially reduces the
amount of heat input needed to explain the observed scaling relations.  
However, radiative cooling unchecked by feedback would lead to overcooling 
of intrahalo gas at $z \sim 2-3$ (see \S~\ref{overcool}).  At least some 
feedback is needed to keep $\sim 90$\% of intergalactic baryons
in gaseous form, as observed.  Mergers following an episode of
feedback can potentially amplify the entropy generated by that
feedback, when low-density gas associated with low-temperature
halos is shocked.  However, the overall impact of mergers on intracluster
entropy remains unclear.  Determining just how cooling, mergers, and
feedback conspire to produce the present-day entropy distribution
of clusters will require high-resolution numerical simulations 
and semi-analytical modeling focusing on the intergalactic entropy 
distribution and how it evolves.

\acknowledgements 

We acknowledge Megan Donahue and Don Horner for helpful
conversations and Trevor Ponman and Stefano Borgani for
comments on the original manuscript.  MLB is supported 
by a PPARC rolling grant for extragalactic astronomy and 
cosmology at the University of Durham.

%\newpage

\end{document}